\documentclass[fleqn,usenatbib]{mnras}
\usepackage{txfonts}
\usepackage[T1]{fontenc}
\usepackage{ae,aecompl}
\usepackage[usenames,dvipsnames]{xcolor}
\usepackage{float}
\usepackage{textgreek}
\usepackage{units}
\usepackage{graphicx}	
\usepackage{caption}
\usepackage{subcaption}
\usepackage{grffile}
\usepackage[nointegrals]{wasysym} 
\usepackage{amssymb}	

\newcommand{\myavg}[1]{$\langle {#1}\rangle$}

\newcommand{\black}[1]{\textcolor{black}{#1}}

\author[Legodi et.\ al.]{L.\ S.\ Legodi$^{1,4}$\thanks{E-mail: slegodi@ska.ac.za}, A.\ R.\ Taylor$^{1,2}$, J.\ M.\ Stil${^3}$\\ 
$^1$ Inter-University Institute for Data Intensive Astronomy, and \\
Department of Astronomy, University of Cape Town, Private Bag X3, Rondebosch 7701, South Africa \\
$^2$ Department of Physics and Astronomy, University of the Western Cape, Private Bag X17, Bellville 7537, South Africa\\
$^3$ Department of Physics and Astronomy, University of Calgary, 2500 University Dr.\,N.W., Calgary, Alberta T2N 1N4, Canada\\
$^4$ South African Radio Astronomy Observatory (SARAO), Black River Park, Observatory, Cape Town, South Africa
}

\title{Broad-band Radio Polarimetry of disk galaxies and AGN with KAT\,7}

\date{Accepted 16 Oct. 2020. Received 25 Nov. 2020; in original form 2019}

\pubyear{2019}

\begin{document}
\label{firstpage}
\pagerange{\pageref{firstpage}--\pageref{lastpage}}
\maketitle

\begin{abstract}
We report broad-band (1.2 - 1.9 GHz) radio continuum observations at arcminute resolutions of two nearby disk galaxies, NGC 1808 and NGC 1097, and four AGN powered radio sources; PKS B$1934-638$, PKS B$0407-658$, J$0240-231$, and J$0538-440$. We use Rotation Measure Synthesis to analyze their Faraday complexity. Observations were made with the KAT\,7 radio telescope array, in South Africa. The AGN powered sources fall into two ``Faraday" categories -- simple and complex. \black{The most polarized sources, J$0538-440$ and J$0240-231$, are found to have complex Faraday spectra that can be time variable (J$0538-440$ case) and also indicative of complex Faraday emitting and rotating components along the line of sight}. PKS B$0407-658$ shows a simple Faraday spectrum while PKS B$1934-638$ is undetected in polarization. The disk galaxies are classified as complex, albeit at low signal-to-noise. \black{This may indicate depolarization due to turbulence of the magnetised plasma in the bar and circumnuclear regions and/or frequency-dependent depolarization at L-band}.
\end{abstract}

\begin{keywords}
radio continuum: galaxies < Resolved and unresolved sources as a function of wavelength, galaxies: disc < Galaxies, galaxies: magnetic fields < Galaxies, polarization < Physical Data and Processes
\end{keywords}


\section{Introduction}\label{intro}
How magnetism fits into the picture of galaxy evolution over cosmic time is a subject of debate among astronomers. The link between magnetic fields and galaxy evolution is theorized \citep[e.g.][]{Lou1998} but 
observational constraints have been lacking, due to technical limitations
\citep[e.g.][]{Chamandy2016}.  
Understanding the evolution of magnetic
fields and their relation to galaxy evolution is one of the key science objectives of next generation radio telescopes such as the Square Kilometre Array
\citep{Johnston-hollitt2015,Heald2020}. The polarization of radio emission is a powerful probe of cosmic magnetic fields.  
\par
Some observational studies in the radio  \citep{Eichendorf1980,Beck2002,Beck2004,Beck2012d,Farnes2014} have provided insights into the workings of magnetism in galaxies.
Highly sensitive and resolved imaging observations of nearby disk galaxies \citep{Beck2005}, show large scale magnetic fields related to spiral structure. 
For more distant galaxies that remain unresolved by current radio imaging telescopes
we must rely on the frequency dependence of polarization, instead of direct imaging, as our primary probe of
magnetic properties \citep{Stil2009a}.  
However, the relatively narrow instantaneous bandwidths typical of most past studies of 
radio polarization limits both the resolution of  Faraday spectral features and the ability 
to characterize the frequency dependent effects of internal depolarization.
\par
Recent studies have begun to probe polarized radio emission over broader bands. For example, \citet{OSullivan2012} presented a broad-band spectropolarimetric study of four
radio bright and highly polarized AGN sources, for which Faraday rotation features could be resolved.
\cite{Anderson2015} report a wide band study of polarimetry in more than 500 sources,
\cite{Anderson2016} observed 36 strongly polarized 
sources over a very wide band from $\sim 1 - 10$ GHz, and find Faraday complexity in most of their sources, asserting that complexity is thus common among strong polarized sources. \black{A more recent study, \citet{Schnitzeler2019}, provides a wide-band (1.3 - 3.1 GHz) polarimetry survey of unresolved (angular resolution $\sim 2^{\prime}\times1^{\prime}$) southern sources. The authors argue in favour of the Galactic foreground being a dominant component in terms of Faraday rotation of background sources, implying simple rather than complex intrinsic distributions of Faraday rotation.}
\par
In this paper we extend the studies of broad-band polarimetry of extragalactic radio sources
by studying a sample of six radio sources, including two disk galaxies and four AGN
dominated sources. 
The study was carried out with observations from the Karoo Array Telescope (KAT\ 7), which was
constructed as an engineering prototype array at the South African SKA site as part of the 
technical development of the MeerKAT SKA precursor telescope. \black{The 64 dish MeerKAT array will form part of the SKA mid-frequency component.}
The KAT 7 wavelength coverage from 1.2 to 1.9 GHz provides us with a \black{good resolution of the Faraday spectrum of polarized emission (see Section \ref{subsec:RMsynth} )}. 
The rotation measure synthesis technique (RM Synthesis), first described by \citet{Burn1966}, is employed to explore the intrinsic polarization properties of the sources.
\par
The spatial resolution limitations of KAT\,7 at low redshift are comparable to those of the more sensitive MeerKAT (and the future full SKA array) at high redshift. MeerKAT has the sensitivity to detect polarization in faint distant objects but does not resolve them, so we explore the peak flux properties of nearby objects to probe the spectropolarimetric behaviour in a fashion that can be extended to more distant and fainter objects with much more powerful instruments. This article is structured as follows: section \ref{Observations+Reductions} discusses our observations and data reductions; section \ref{PolarimetrySect} presents the results of our polarimetric analysis; \black{we discuss our findings in section \ref{Discussion}}; and we summarise, discuss and also  make conclusions based on this work in section \ref{Summary+conclusions}. 


\section{Observations and Data Reduction}\label{Observations+Reductions}

\subsection{Sample Selection}

We selected our sample of disk galaxies from the source list of the MeerKAT HI 
Observations of Nearby Galactic Objects: Observing Southern Emitters survey\footnote{PI: Erwin de Blok, \url{http://mhongoose.astron.nl/}}, ``MHONGOOSE" \citep{MHONGOOSE}. The survey will secure very high sensitivity atomic hydrogen images of a sample of nearby galaxies. A further criterion was that our sources be sufficiently radio bright, with a reported integrated total intensity \black{greater than 200 mJy at L band. The flux limit was chosen so as to have sources with sufficient flux density to be detectable by KAT\,7 at a few percent polarization}. This resulted in a sample of two disk galaxies.  We also observed four AGN dominated sources, with three of them serving as calibrators for the remaining targets. \black{The two disk galaxies both have unique features in the context of the overall disk galaxy population, but the analysis done here will shed some light on the polarization resulting from the large scale magnetic field in disk galaxies that are unresolved.}

\subsubsection*{Disk Galaxies} \label{subsubsec:disk gals}
{\bf NGC 1097} is identified as a Sy1 galaxy \citep{StorchiBergmann1997} with a circumnuclear star-burst ring \citep{Gerin1988}. It is a barred spiral, SBb, with its radio emission dominated by the circumnuclear ring and nucleus \citep{Beck2005}. The magnetic properties of NGC 1097 have been studied  using narrow band ($\sim 50$ MHz) high resolution radio data ($2^{\prime\prime} - 15^{\prime\prime}$ resolutions) at 4.8 and 8.4 GHz, with the NRAO Very Large Array, VLA \citep{Beck2002,Beck2005}. The galactic magnetic field was observed to control the ISM gas flow at kpc scales and found to have a regular component that follows the shape of the spiral arms \citep{Beck2005}. The galaxy has a redshift of 0.0042.
\par
{\bf NGC 1808} is identified as a Sy2 galaxy \citep{Tacconi-Garman1996,Galliano2008} with a low luminosity  AGN \citep{Veron-Cetty1985,Dahlem1990,Gonzalez-Martin2013,Esquej2014}. It was classified as a peculiar Sbc galaxy \citep{Veron-Cetty1985}. It also has a star-burst region in the inner $\sim 750$ pc dominating the infra-red emission \citep{Sengupta2006}. From observations at 6 and 20\,cm wavelengths with the VLA, \citet{Dahlem1990} found linear polarization degrees of up to $30\%$ and a steep ($\alpha=-0.88$) radio spectrum from outside the central region of the galaxy. The galaxy has a redshift of 0.0033.

\subsubsection*{AGN}
{\bf J0240$-$231} has been identified to be a QSO and also a Gigahertz-peaked spectrum (GPS) source \citep{Kuehr1981,ODea1998,Veron-Cetty2006}. It has a redshift of z = 2.22 \citep{Hewitt1989}. The source has been observed to show significant linear polarization at 4.8 GHz where $p = 3.47 \%$ \citep{Edwards2004}. VLBI studies show structure resembling two lobes that are separated by 12 mas \citep{Dallacasa1998}.
\par
{\bf J0538$-$440} is reported to be a BL Lac, GPS object and also a variable $\gamma$-ray source located at redshift 0.896 \citep{Impey1988,Romero2000,Tornikoski2001,Romero2002,Andruchow2005,Torniainen2005}. It has also been identified as a potential gravitational lens \citep{Surpi1996}. \black{This source presents a time variable total intensity spectrum, at cm and mm wavelengths, as can be seen in calibrator databases such as the ATCA Calibrator Database\footnote{\url{https://www.narrabri.atnf.csiro.au/calibrators/calibrator_database_viewcal.html?source=0537-441} }. It has also been observed to be highly polarized and variable in the optical \citep{Impey1988,Romero2000,Romero2002,Andruchow2005}.}
\par
\textbf{PKS B$1934-638$} is a quasar and GPS source  at redshift 0.18 while \textbf{PKS B$0407-658$} is a quasar and possible compact steep-spectrum (CSS) source \citep{ODea1991,Labiano2006} at a redshift of 0.96. VLBI imaging of PKS B$1934-638$ shows that it has a double lobed structure \citep{Tzioumis1989}. \black{Henceforth, we use the name "PKS $1934-638$" for PKS B$1934-638$ in this work as this name is most commonly used for the source.}

\begin{table*}
\caption{Summary of the KAT\,7 observations. The table lists the radio sources (some aliases in brackets) observed during each run, their J2000 coordinates, the observation dates (and times given in UTC), central frequency ($\nu_{c}$), number of antennas available ($N_{\rm ant}$), and the total recovered bandwidth ($\Delta \nu$).}
\begin{tabular}{lllcccc}
\hline\hline\\
[-1.25ex]
Source (aliases)  & RA (J2000) & DEC (J2000) & Date (UTC time) & $\nu_{c}$  & $N_{\rm ant}$ & $\Delta \nu$ \\ 
& HH:MM:SS & DD:MM:SS &&(GHz)&& (MHz) \\
[1.25ex]
\hline  
NGC 1808 & 05:07:42.3000 & -37:30:47.0000 & 29/07/2016 (01:07:20.1 - 07:12:30.4) & 1.894 & 6 & 152 \\ 
PKS $1934-638$ (J1939-6342) & 19:39:25.0267 & -63:42:45.6255 & &  & &  \\ 
J$0538-440$ (J0538-4405, 0537-441) & 05:38:50.1800 & -44:05:10.3000  & &  & &  \\ 
PKS B$0407-658$ (J0408-6545) & 04:08:20.3788 &-65:45:09.0806 & &  & &  \\ 
-- & -- & -- & -- & -- & -- & --\\ 
&&&&&&\\ 
NGC 1808 &&& 31/07/2016 (01:00:19.5 - 07:03:29.8) & 1.394 & 6 & 148 \\ 
PKS $1934-638$ &&&& & & \\ 
J$0538-440$ &&&& & & \\ 
PKS B$0407-658$ &&&& & & \\ 
-- & -- & -- & -- & -- & -- & --\\ 
&&&&&&\\ 
NGC 1097 &02:46:19.0000 & -30:16:30.0000 & 09/06/2015 (02:32:11.9 - 08:34:42.2) & 1.850 & 7 & 223 \\ 
PKS $1934-638$ &&&& & & \\ 
J$0240-231$ (J0240-2309, 0237-233) & 02:40:08.1400 & -23:09:16.0000 && & & \\ 
PKS B$0407-658$ &&&& & & \\ 
&&&&&&\\ 
-- & -- & -- & -- & -- & -- & --\\ 
NGC 1097 &&& 15/06/2015 (02:30:22.9 - 08:35:13.2) & 1.350 & 7 & 211 \\ 
PKS $1934-638$ &&&& & & \\ 
J$0240-231$ &&&& & & \\ 
PKS B$0407-658$ &&&& & & \\ 

\hline 
\end{tabular} 
\label{Obstable1}
\end{table*}

\subsection{Observations}

KAT\,7 is a synthesis array telescope located adjacent to the 
MeerKAT and SKA phase 1 central site in  the Karoo plateau of South Africa.  
It was built as an engineering test-bed preceding the 64 antenna SKA mid-frequency precursor array, MeerKAT. KAT\,7 consists of seven 12-metre, centre-fed parabolic antennas in a fixed configuration with baselines ranging from  26  to 185 metres \citep{Carignan2013}. 
It has a maximum instantaneous bandwidth of 256 MHz within the RF range 1200 - 1950 MHz.   
Two components (X and Y) of linear polarization are detected with prime-focus feeds on each antenna.
We observed in wide-band, full-polarization mode, providing raw data consisting of four polarization correlation products sampling a
256 MHz band in 1024 spectral channels, each of width 390.625 kHz.
Our observations covered the entire 750\,MHz RF band using three separate observing runs for each disk galaxy target with the 256 MHz centred near 1350, 1600, and 1850 MHz (see Table \ref{Obstable1}).
\par
Each observation consisted of 12 hour tracks during which on-source observations of 4 minutes of either NGC\,1097 or NGC\,1808 were alternated with 2 minute scans of a phase calibrator, either J$0240-231$ (for NGC\,1097) or J$0538-440$ (for NGC\,1808). Flux and bandpass were calibrated with several 3 minute scans on PKS $1934-638$ and PKS B$0407-658$ depending on which calibrator was visible. The cadence on these flux calibrator scans was approximately an hour. The 12 hour tracks provided a good range of parallactic angle variation for polarization calibration.

\subsection{Data reduction} \label{Calibration and Imaging}

Processing of the raw visibility data was carried out using the Common Astronomy Software Applications, {\sc casa} \citep{Mcmullin2007} package. After removal of the channels at the upper and lower ends of the band due to roll-off of the bandpass shape, the visibility data were inspected visually and manually flagged to remove radio frequency interference (RFI) signals. The mid-band, centred at 1600 MHz, was heavily populated with RFI. Analysis in this band was severely compromised as a result and we did not include the mid-band in our subsequent analysis. 
The objects observed, recovered band pass, and number of operating antennas for each observing run are listed in Table~\ref{Obstable1}.
\par
PKS $1934-638$ was used for absolute flux calibration, and to measure the complex bandpass shape and 
cross-hand delay for all observations. The sources J$0538-440$ and J$0240-231$ were used as secondary calibrators to track time-dependent amplitude and phase calibration over the course of each observation and for on-axis polarization leakage calibration given the large parallactic angle covered.
The time-dependent total intensity gain is coupled to any detectable 
polarization signal from the astronomical sources and telescope. The broad parallactic angle coverage of the observations of the secondary calibrators allows us to solve for the polarization of the calibrators from the initial antenna based time-dependent gain solutions as well as the frequency-dependent XY-phase calibration.  The source polarization solutions are then used to correct the time-dependent gains and to solve for frequency-dependent polarization leakage. {\sc casa} polarization calibration tasks do not account for rotation measure across a wide band and thus the polarization solution from the gains is the average across the band. This is not ideal for a non-zero rotation measure. This effect is not mitigated during data reduction in this work and thus may affect polarization measurements. We assume that the effect will not be significant at the bandwidths of our data, $\sim 200$ MHz, and that none of the targets have large RM values. For example, \cite{Taylor2009} report an RM of $14.5\pm 2.2$ rad\,m$^{-2}$ with polarization degree of $0.8 +/- 0.1\%$ at 1.4 GHz for J$0240-231$.
\par
We do not observe a source with known absolute polarization position angle.  Therefore, while the relative polarization position angles (with respect to frequency) are calibrated, the absolute polarization position angles for the polarization solutions are not known. The instrumental polarization is accounted for by solving for leakages using a source with a large parallactic angle coverage \citep[e.g.][]{Jagannathan2017}.
\par
After applying the calibration solutions, Stokes  $I$, $Q$, and $U$ image cubes were made for each source. The imaging and deconvolution was done using the Clarke-Stokes algorithm implemented in {\sc casa}'s {\sc tclean} task. \black{We averaged the visibility channels into 3.9 MHz channels to construct the image cubes which resulted in the maximum detectable Faraday depth decreasing by an order of magnitude (see Section \ref{subsec:RMsynth} and Table \ref{RMresolution}).} The typical synthesized beam at the low band is $\sim 4^{\prime}.9$ and $\sim 3^{\prime}.6$ at the high band. The observed angular size of our sources is consistent with unresolved sources.

\begin{table*}
\caption{Band averaged values of Stokes \myavg{I}$_{\nu_c}$, band averaged linear polarization intensities (\myavg{P}$_{\nu_c}$ and bias corrected \myavg{P_o}$_{\nu_c}$), and also Q, U noise ($\sigma_{QU}$). The error in the averaged values are $\sigma/\sqrt{N}$, where $N$ is the number of channels, $\sigma$ is the standard deviation of the values and $\nu_c$ is the frequency associated with the band. The low and high bands are indicated by sub-/superscripts $1.35$ and $1.85$, respectively indicating $\nu_c$ in GHz. All values of intensity have units of mJy per beam, abbreviated as mJy/bm. The last column displays the mean spectral index, -\myavg{\alpha}$_{1.85}^{1.35}$, over the two KAT 7 bands} 
\begin{tabular}{lrrrcrrrcc}
\hline \hline\\
[-1.25ex]
Source & \myavg{I}$_{1.35}$\ \ \ \hfill & \myavg{P}$_{1.35}$\ \  \hfill & \myavg{P_o}$_{1.35}$\  \hfil & \myavg{\sigma}$_{QU}^{1.35}$\hfill & \myavg{I}$_{1.85}$\ \ \ \hfill & \myavg{P}$_{1.85}$\ \ \hfill & \myavg{P_o}$_{1.85}$\ \hfill & \myavg{\sigma}$_{QU}^{1.85}$ & $\alpha_{1.85}^{1.35}$\\ 
 & ($\mathrm{\frac{mJy}{bm}}$)\ \ \ \ & ($\mathrm{\frac{mJy}{bm}}$)\ \ \ & ($\mathrm{\frac{mJy}{bm}}$)\ \ \ & ($\mathrm{\frac{mJy}{bm}}$) & ($\mathrm{\frac{mJy}{bm}}$)\ \ \ \ & ($\mathrm{\frac{mJy}{bm}}$)\ \ \ & ($\mathrm{\frac{mJy}{bm}}$)\ \ \ & ($\mathrm{\frac{mJy}{bm}}$)& \\
[1.25ex]
\hline 
NGC 1808    & $514.6 \pm 0.2$ & $2.4\pm 0.5$ & $2.4\pm 0.5$ & $0.3$& $400.3 \pm 0.2$ & $2.2\pm 0.4$ & $2.2\pm 0.4$ &$0.3$ &$-0.82\pm0.01$\\ 
NGC 1097    & $340.8 \pm 0.3$ & $2.6\pm 0.4$ & $2.6\pm 0.4$ & $0.4$& $248.0 \pm 0.1 $& $3.6\pm 0.3$ & $3.5\pm 0.3$ &$0.2$ &$-1.06\pm0.02$\\ 
J$0240-231$  &  $5727.4\pm1.2$ & $58.9\pm 3.6$& $58.9\pm 3.6$& $0.6$& $4861.1\pm 0.8$ &$101.5\pm 3.1$&$101.5\pm3.1$ &$0.5$ &$-0.56\pm0.01$\\ 
J$0538-440$  &  $3923.0\pm 0.3$& $27.6\pm 2.8$& $27.6\pm 2.8$& $0.6$ &$3806.5\pm 1.3$ & $10.9\pm 2.5$& $10.9\pm2.5$ &$0.4$ &$-0.10\pm0.01$\\ 
PKS $1934-638$& $14876.9\pm 0.5$& -- & -- & $1.5$ &$13315.6\pm 0.3$& -- & -- &$1.0$ &$-0.36\pm0.01$\\ 
PKS B$0407-658$&$15130.7\pm 1.0$& $18.7\pm 9.6$& $18.7\pm 9.6$&$1.5$ & $10848.2\pm 1.0$&$21.7\pm 6.7$ &$21.7\pm 6.7$ &$1.3$ &$-1.12\pm0.01$\\ 
\hline 
\end{tabular} 
\label{StokesI+pTable}
\end{table*}

\begin{figure*} 
\includegraphics[width=\textwidth]{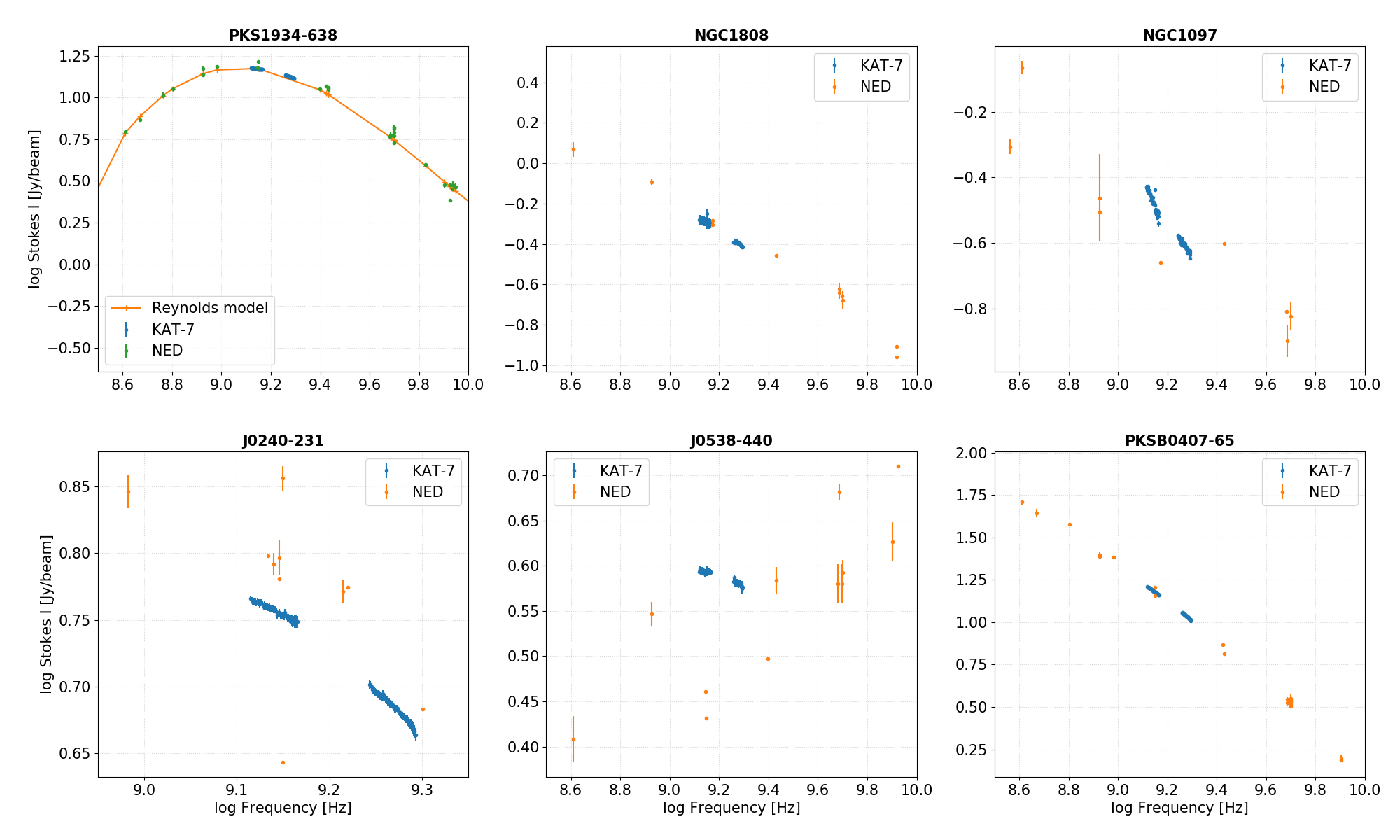}
\caption{Total intensity across KAT-7 band (blue points), some corresponding literature values (from NED, \url{https://ned.ipac.caltech.edu/}) at frequencies near the KAT-7 band. \textbf{Top panels}: \black{Left: The primary flux calibrator, PKS $1934-638$, with the Reynolds 1994 model plotted as a red curve \citep{Reynolds1994}}. Middle: NGC 1808. Right: NGC 1097. \textbf{Bottom panels}: Left: The polarization calibrators, J$0538-440$ and J$0240-231$ (middle). Right: PKS B$0407-658$.}
\label{Tot_flux_plots}
\end{figure*}

\begin{figure*} 
    \begin{subfigure}{1.0\hsize}
    \includegraphics[width=\textwidth]{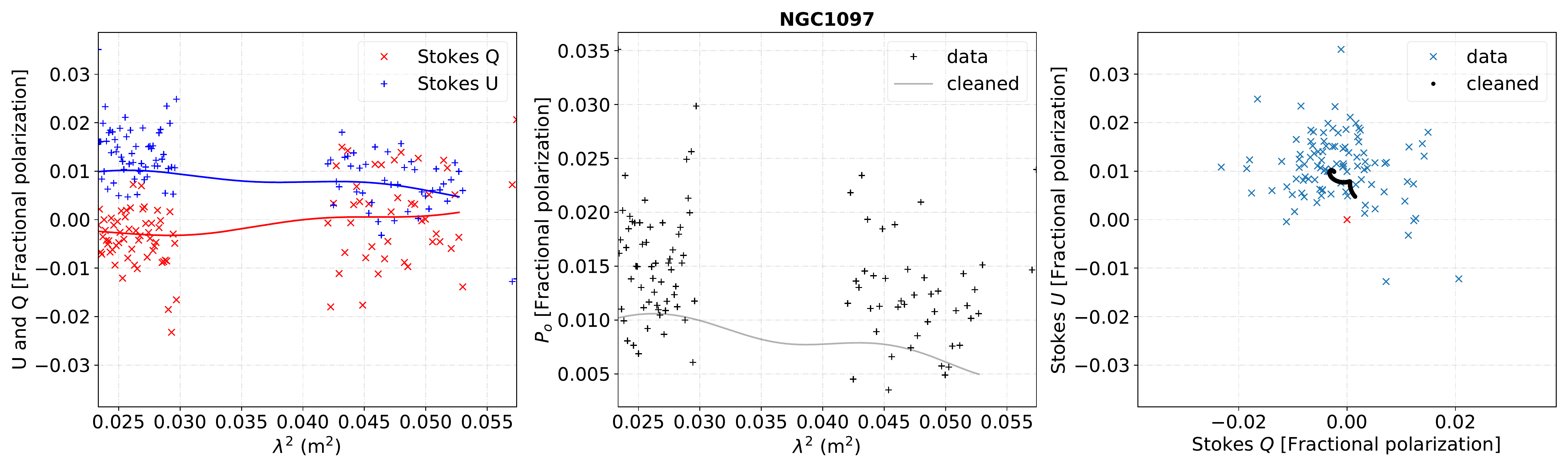}\\
    \includegraphics[width=\textwidth]{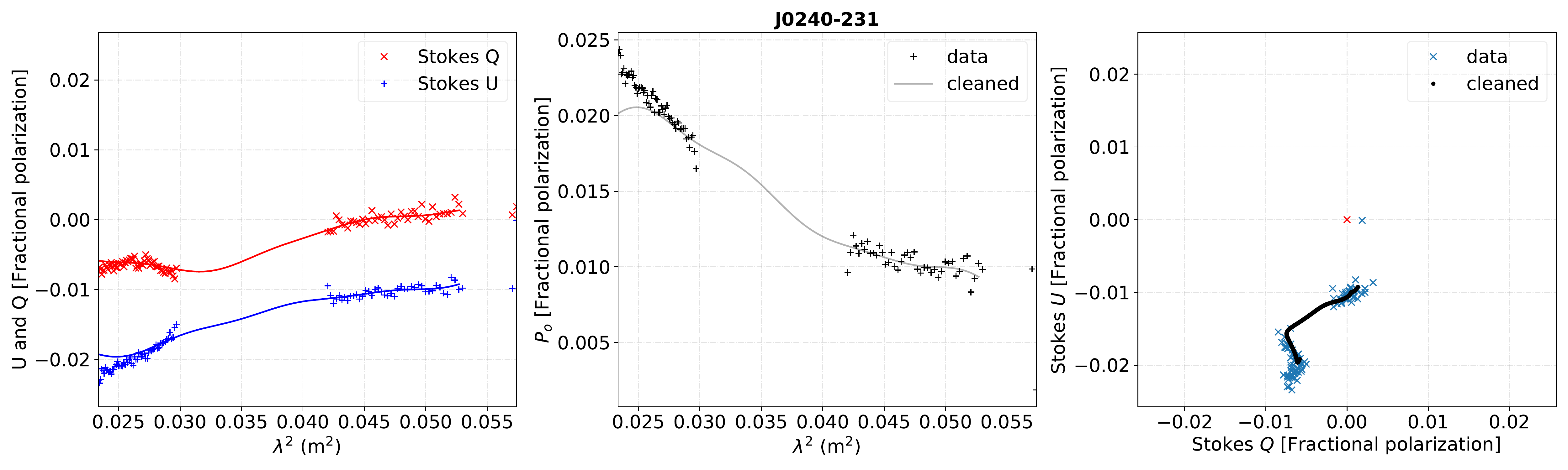}
    \subcaption{\textbf{Top panels}: $q(\lambda^2)$ and $u(\lambda^2)$, $p(\lambda^2)$, and $q$ vs. $u$ for NGC 1097. \textbf{Bottom panels}: Same measurements for J$0240-231$. See main caption below.}
    \label{pquX_fullband-ngc1097}
    \end{subfigure}

    \begin{subfigure}{1.0\hsize}
    \includegraphics[width=\textwidth]{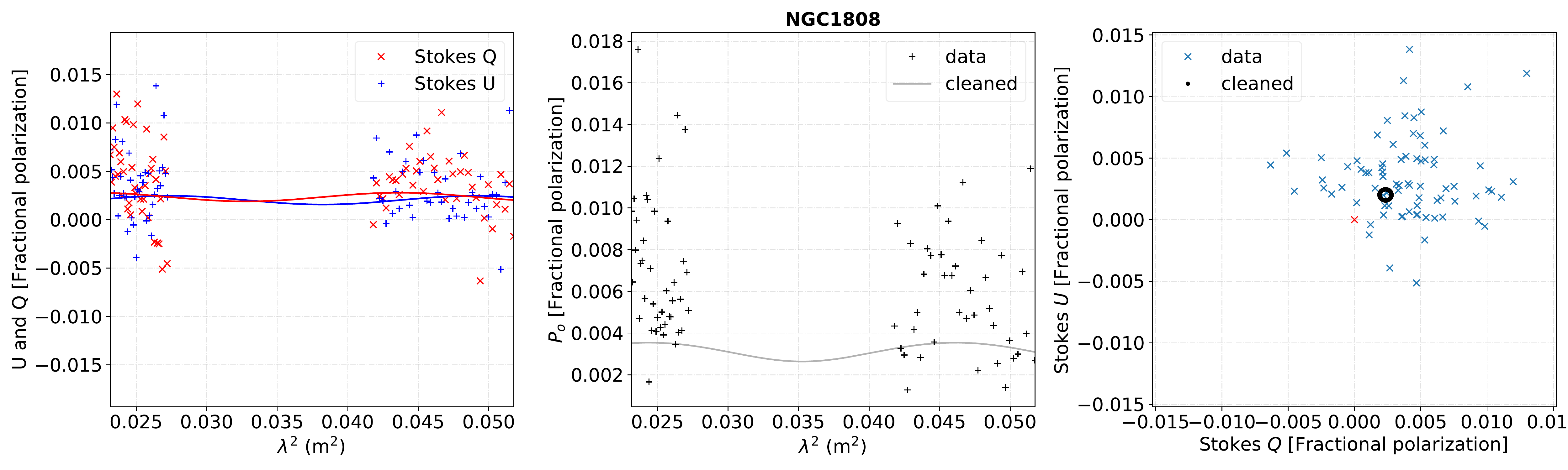}\\
    \includegraphics[width=\textwidth]{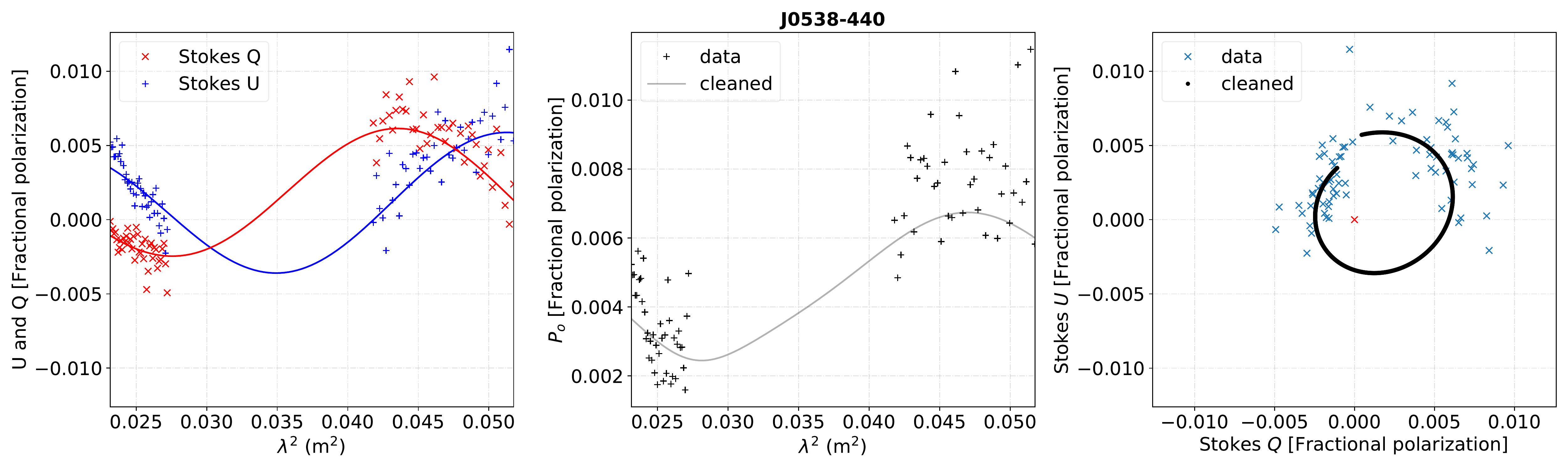}
    \subcaption{Same as in Figure \ref{pquX_fullband-ngc1097}, \textbf{top panels}: Measurements for NGC 1808. \textbf{Bottom panels}: Same measurements for J$0538-440$. See main caption below.}
    \label{pquX_fullband-ngc1808}
    \end{subfigure}
\end{figure*}

\begin{figure*}\ContinuedFloat
    \begin{subfigure}{1.0\hsize}
    \includegraphics[width=\textwidth]{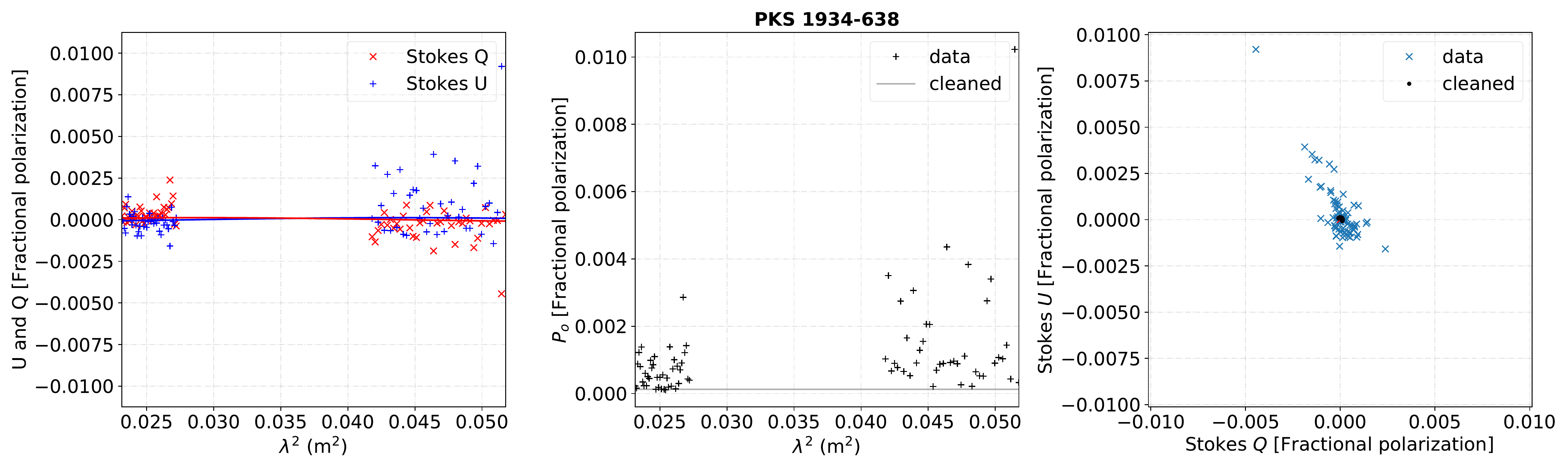}\\
    \includegraphics[width=\textwidth]{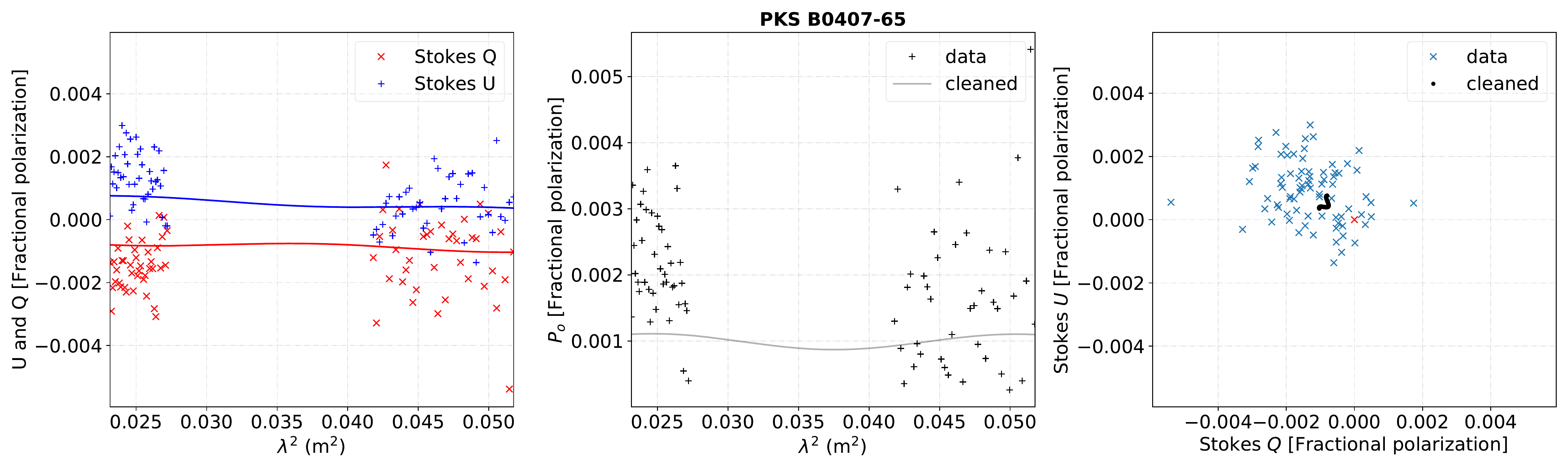}
    \subcaption{Same as in Figure \ref{pquX_fullband-ngc1097}, \textbf{top panels}: PKS $1934-638$. \textbf{Bottom panels}: Same measurements for PKS B$0407-658$. See main caption below.}
    \label{pquX_fullband-fluxcals}
    \end{subfigure}
\caption{Figures \ref{pquX_fullband-ngc1097} - \ref{pquX_fullband-fluxcals}: Polarization measurements, fractional Stokes $Q$ and $U$ \black{(with errors, error bars omitted, of the order $\sim 10^{-5}$)} and also total linear polarization $p_0$, across the KAT-7 band along with the RM ``cleaned" versions of each quantity -- in each panel, shown as curves in the leading two plots and black points in the trailing plot.}
\end{figure*}

\section{Results} 
\label{PolarimetrySect}

\subsection{Spectropolarimetry}

We extracted spectra in Stokes $I$, $Q$ and $U$ for each source at the peak total intensity 
in each cube channel. The errors for each of the peak intensities are estimated by taking the standard deviation of intensities at source free/off-source regions in each cube channel. We then propagate these errors for derived quantities, such as linear polarization, calculated from $I$, $Q$ and $U$. The peak intensity values are measures of the integrated intensity, as these sources are unresolved.
\par
Per-channel Stokes $I$ intensities for each source are shown as blue dots in each panel of Figure \ref{Tot_flux_plots}.  
For comparison we also show flux densities for each source over a broader frequency range obtained from the NED database.   
There is very good agreement with previously reported values for PKS $1934-638$, 
NGC 1808 and PKS B$0407-658$. The spectrum, for NGC 1097, while consistently steep, is very poorly defined by previous data.  
However, our data fall within the  general trend of past observations. 
\black{The data for J$0538-440$ suggest variability. Our measured spectrum differs significantly from the literature values. The discrepancy is brought about most likely by time variability in the  J$0538-440$ spectrum as can be seen on the ATCA database}. The source has a total intensity spectrum that is flat near $\sim 1$ GHz with a standard deviation of $\sim 1.0$ Jy in the region while showing a bump towards, and also reaching a peak near, $\sim 10$ GHz. The results for the flat spectrum source J$0240-231$ also indicate variability.
The total flux of J$0240-231$ in the 1.36 - 2.0 GHz frequency range  has been reported to vary over the range 4.4 - 7.1 Jy 
\citep{Kuehr1981, Condon1998, Stanghellini1998a, Tingay2003a, Stanghellini2005}.

\begin{table*}
\centering
\caption{Degrees of linear polarization (in percentage) from \citet{Eichendorf1980} along with corresponding values form the KAT-7 bands in this work; \myavg{p_o}$_{1.35}$ and \myavg{p_o}$_{1.85}$. Subscripts are as in Table \ref{StokesI+pTable}. }
\begin{tabular}{lllllll}
\hline\hline\\
[-1.25ex]
Source & $p_{5.00}$ & $p_{2.73}$ & $p_{1.67}$& $p_{1.43}$ & \myavg{p_o}$_{1.35}$& \myavg{p_o}$_{1.85}$  \\
 & (\%) & (\%)  &  (\%)  &  (\%) &  (\%) & (\%) \\
[1.25ex]
\hline
NGC 1808 & -- & -- & -- &          -- & $0.5\pm0.09$ & $0.5\pm0.10$\\
NGC 1097 & -- & -- & -- &          -- & $0.8\pm0.11$ & $1.4\pm0.11$\\
PKS $1934-638$  & 0.1 & 0.3 & 0.4 & 0.2 & -- & --\\
PKS B$0407-658$ & 0.5 & 0.4 & 0.6 & 0.4 & $0.2\pm0.07$ & $0.2\pm0.07$\\
J$0240-231$    & 5.4 & 3.4 & 1.7 & 1.8 & $1.0\pm0.07$ & $2.1\pm0.07$ \\
J$0538-440$ & -- & -- & -- &        -- & $0.7\pm0.08$ & $0.3\pm0.07$\\
\hline 
\end{tabular}
\label{Eichendorf_1980Table}
\end{table*}

We derive the bias-corrected \citep{Simmons1985} linear polarized flux density  as 

\begin{equation}
P_{\rm o} = \sqrt{P^2- \sigma_{QU}^2}, \\
\end{equation}
where $P^2=Q^2+U^2$ and $\sigma_{QU}$ is the noise in the peak $Q$ and $U$ values derived as the rms of the amplitudes of off-source positions in each channel. For channel-averaged quantities \myavg{Q} and \myavg{U}, the noise is reduced by $1/\sqrt{N}$, where $N$ is the number of channels. Band averaged polarization is given by \myavg{P_o}. Similarly, the band averaged fractional polarization is given by 

\begin{equation}
\langle p_o\rangle = \frac{\langle P_o \rangle}{\langle I_{pow}\rangle}.
\end{equation}
with fractional $Q$ and $U$ given by $u = U/I_{pow}$ and $q = Q/I_{pow}$. We compare band-averaged polarization values with previous studies. $I_{pow}$ is the simple power law fit to the measured peak total intensity: 
\begin{equation}
I_{pow}= I_o\big(\frac{\nu}{\nu_o}\big)^{\alpha}. 
\end{equation}
The bias correction method applied best approximates $p_{\rm o}$ when $p/\sigma_{qu}\gtrsim 3$.  Here $\sigma_{qu}$ is the rms noise in $q$ and $u$. Table~\ref{StokesI+pTable} summarizes the band averaged total and polarized flux densities we derived for each source in each band, as well as the total intensity spectral index between the two bands.
\par
The recent study by \cite{Jagannathan2017} characterized the effect of instrumental polarization leakage in interferometric imaging. Such effects are much reduced for long integrations with large coverage in parallactic angle and also for sources nearer to the pointing centre of the observed field. We use the secondary calibrators to calibrate on-axis leakages. We find leakage terms at the level of $\sim 0.06\%$. PKS $1934-638$ was observed during each observing run as the primary calibrator. \black{This source has no detectable polarization at L-band and is a well known Southern hemisphere flux calibrator \citep[e.g.][]{Reynolds1994}.} Our average observed percent polarization for PKS $1934-638$ from the two frequency bands is $p = 0.06 \pm 0.07$ \%, indistinguishable from the on-axis instrumental polarization. The most conservative estimate of instrumental polarization then, is to assume that PKS $1934-638$ is unpolarized at this level of sensitivity and that the observed polarization is entirely instrumental. We have thus included in Table \ref{StokesI+pTable} an additional uncertainty of $0.06 \%$ in quadrature to the formal noise derived error on the linear polarization degree. 
\par
The fractional polarization degree of each source in each band are listed in Table \ref{Eichendorf_1980Table}.
The table also lists some polarization fractions at similar frequencies found in the literature from \citet{Eichendorf1980}, who presented a 
collection of polarization properties (position angles, flux densities, and rotation measures) that were measured 
between 1965 and the middle of 1974 and these include data at 5.00, 2.73, 1.67, and 1.43 GHz -- listed in Table~\ref{Eichendorf_1980Table}. The typical observational errors in their data are in the range of 1 to 2 percent in 1965 while being better at 0.5 to 1 percent in 1974. Their data set contains observations for only three of our sources. \\
Linear polarization for just one of our targets, NGC 1097, was measured \citep{Beck2002,Beck2005} at 22, 18, 6, and 3 cm (corresponding to $\sim 1.36,\ 1.66,\ 5.00 $ and $\sim 10.00$ GHz, respectively) and also at 4.80 GHz \citep{Stil2009a}. The integrated linear polarization degrees for NGC 1097 were reported at low significance where $p_{\rm 22cm}=1.5\pm0.9\%$ and $p_{\rm 18cm}=1.3\pm1.0\%$ at resolutions of $2^{\prime\prime}-15^{\prime\prime}$, but still consistent with our results calculated from the peak polarized flux: $p_{\rm 22cm}=0.8\pm0.1\%$ and $p_{\rm 16cm}=1.4\pm 0.1\%$ (see Table \ref{Eichendorf_1980Table}). The source was undetected at 4.80 GHz where $p_o = 0.7\pm0.8\%$.

\subsection{RM Synthesis} \label{subsec:RMsynth}
The electric field vector of linearly polarized electro-magnetic waves is rotated through the Faraday process, as the waves pass through a magneto-ionic medium. The component of the magnetic field along our line of sight (LoS) can be probed through observations of this emission. \cite{Burn1966} presented a technique that can resolve the different components of magnetic fields that alter the polarization of the incident radiation along the LoS. The technique was later expanded and dubbed ``Rotation Measure Synthesis" (RM-Synthesis, for short) by \cite{Brentjens2005}. 
\par
We apply the Rotation Measure Synthesis technique to analyze the intrinsic polarization properties of our sources. The technique transforms $q(\lambda^2)$ and $u(\lambda^2)$ to $q(\phi)$ and $u(\phi)$ through a Fourier transform  -- $\phi$ is the Faraday depth, the line of sight integral of the thermal electron number density (denoted as $n_e$) and magnetic field ($\vec{B}$) component along the observer's line of sight: 

\begin{equation}
\phi = K_c \int^0_r n_e\vec{B}\cdot d\vec{r}. 
\end{equation}
$K_c$ is a constant. Faraday rotation is quantified by the Rotation Measure (or RM) in units of rad\,m$^{-2}$. The RM is defined by the change in polarization angle $\chi(\lambda^2)$ as a function of the square of the wavelength, $\lambda^2$:

\begin{equation}
{\rm RM} = \frac{d\chi(\lambda^2)}{\ d\lambda^2},
\end{equation}
with the polarization angle given by

\begin{equation}\label{eq:Xi-linear}
\chi = \frac{1}{2}\arctan\frac{u}{q}.
\end{equation}
The Faraday depth and the rotation measure are the same in the simplest case where there is only a single source of non-varying Faraday rotation. This is the so called Faraday simple case \citep[e.g.][]{Brown2017}, where polarization angle, $\chi$, is related to the Faraday depth by 

\begin{equation}\label{eq:PA}
\chi=\chi_0 + {\rm RM}\lambda^2.
\label{eq:simpleFaraday}
\end{equation}
Equation \ref{eq:simpleFaraday} may be valid at modest bandwidths, where $\chi(\lambda^2)$ is linear, but this is not a guarantee. 
In the case that the Faraday rotation plasma and the emitting plasma are mixed, this simple relationship cannot be assumed  \citep{Farnsworth2011}. Observations over broad bandwidths can
display more complex Faraday rotation. 
To characterize such Faraday complexity we use the Faraday spectrum, denoted as $F(\phi)$ \citep{Sun2015}, referred to as the Faraday dispersion function \citep{Brentjens2005}. Such spectra can reveal 
Faraday components in polarized emission at different Faraday depths, $\phi$. $F(\phi)$ is related to linear polarized intensity, $P$ by:
 
\begin{eqnarray}
P(\lambda^2)= \int^{\infty}_{-\infty}F(\phi)e^{2i\phi\lambda^2}d\phi .
\label{PFtransform}
\end{eqnarray}
%
\begin{table*}
\centering
\caption{Key parameters in our RM Synthesis analysis. f$_{c}$ is the frequency range covered, ``Target'' refers to the disk galaxy target's observation run from which the quantities in the table are derived, $\delta\lambda^2$ is the spacing in $\lambda^2$, $\lambda^2_{min}$ and $\lambda^2_{max}$ are the minimum and maximum values, respectively, of $\lambda^2$ corresponding to each band. $\phi_{max-scale}$ is the largest Faraday depth scale to which our RM Synthesis analysis is sensitive and $\vert\phi_{max}\vert$ is the largest depth we can detect. $\delta\phi^{\prime}$ is the RMTF FWHM due to the non-continuous $\lambda^2$ coverage in this work while $\delta\phi$ is the same for the continuous case.}
\footnotesize{\begin{tabular}{ccccccccc}
\hline\hline\\
[-1.25ex]
f$_{c}$ & Target & $\delta\lambda^2$ & $\lambda^2_{min}$ & $\lambda^2_{max}$ & $\phi_{max-scale}$ & $\delta\phi$ & $\delta\phi^{\prime}$ & $|\phi_{max}|$ \\
 & & [m$^2$]  & [m$^2$]  & [m$^2$] & [rad\,m$^{-2}$]& [rad\,m$^{-2}$] & [rad\,m$^{-2}$] & [rad\,m$^{-2}$]\\
[1.25ex] 
\hline\\
[-1.25ex] 
1.251 - 1.962 GHz & NGC 1097 & $9.29\times10^{-5}$ & $2.33\times10^{-2}$ & $5.74\times10^{-2}$ & 134.6 & 53.1 & $101.0\pm 0.8$ & $1.86\times10^{4}$\\
1.318 - 1.970 GHz & NGC 1808 & $9.18\times10^{-5}$ & $2.32\times10^{-2}$ & $5.17\times10^{-2}$ & 135.6 & 67.7 & $98.0\pm 0.8$  & $1.89\times10^{4}$\\
\hline
\end{tabular}}
\label{RMresolution}
\end{table*}

To perform RM Synthesis, we used a grid of trial $\phi$ values, spaced by 0.1 rad/m$^2$ ($\phi \in [-1000, 1000]$ rad\,m$^{-2}$), to measure $F(\phi)$. The inverse Fourier transform to derive $F(\phi)$ is approximated by a summation as observations do not cover a continuous wavelength range and also $\lambda > 0$ at all times:

\begin{equation}
F(\phi) \approx K\sum_{k} P(\lambda_{k}^2)e^{-2i\phi(\lambda^{2}_{k}-\lambda^{2}_{0})}
\label{eq:FsumAprox2}
\end{equation}

$K$ is the inverse of the $\lambda^2$ integral of the sampling function, $W(\lambda^2)$ (which is zero everywhere beyond the limits of the observed $\lambda^2$ range but non-zero where the range is sampled):

\begin{equation}
K = \left(\int^{\infty}_{-\infty} W(\lambda^2)\ d\lambda^2\right)^{-1},
\end{equation}

We progress through the following steps in applying RM Synthesis and deconvolving the intrinsic Faraday spectrum of each source:

\begin{itemize}
\item[1.] Calculate the Faraday spectrum for an input $q, u$ array to avoid the effects of a varying spectral index across a broad band which affects $Q$ and $U$.

\item[2.] Perform a deconvolution on the Faraday spectrum (i.e perform an ``RM clean" \citep{Heald2009} with a loop gain of 0.1. This deconvolution is iterated in a computational loop set to terminate at the RM clean threshold, explained below. \black{The RM synthesis analysis is done in the range $-1000$ to $+1000$ rad m$^{-2}$. RM ranges up to $\vert\phi_{max}\vert$ (see Table \ref{RMresolution}) were also tested and no emission components were detected beyond $\pm 1000$ rad m$^{-2}$.}

\item[3.] The result of the RM clean loop is then a multi-component model with each $\phi$-component characterized by a position angle, peak amplitude of the Faraday spectrum ($|F(\phi)|$), and a $\phi$ value.

\item[4.] We then create RM-cleaned $F(\phi)$, $q(\phi)$, $u(\phi)$, $q(\lambda^2)$, and also $u(\lambda^2)$ data sets based on the multi-component model.
\end{itemize}

Table \ref{RMresolution} lists the approximate values of parameters quantifying the limits of our KAT\,7 data with regard to RM Synthesis. The predicted resolution in Faraday space, $\delta\phi$, is slightly worse (67.7 rad\,m$^{-2}$) in the observations of NGC 1808, J$0538-440$, PKS B$0407-658$, and PKS $1934-638$ than for NGC 1097 and J$0240-231$ (where $\delta\phi = 53.1$ rad\,m$^{-2}$) due to a larger fraction of the band being flagged in the former case. \black{The $\lambda^2$ sampling of KAT 7 allows detection of RM components up to $\vert\phi_{max}\vert \sim 1.86 - 1.89\times10^{4} $ rad\,m$^{-2}$, where $\vert\phi_{max}\vert \approx \sqrt{3}/\delta\lambda^2$, \citep{Brentjens2005}}. \black{$\vert\phi_{max}\vert$ is calculated at the $3.9$ MHz channel width where $\delta\lambda^2\sim 9\times10^{-5}$ m$^2$ as listed in Table \ref{RMresolution}. The worst-case unaccounted rotation (occurring at the lowest wavelength where $\lambda^2_{max} \approx 0.05$ m$^{2}$)  would therefore be $\sim 196^{\circ}$}. Note that the values listed in Table~\ref{RMresolution} are strictly correct for only continuous frequency coverage between the high and low frequencies of the
observations. The effect of the absence of the central band is to create strong ``side-lobes'' of the ``dirty'' Faraday transfer function. The quantity $\delta\phi^{\prime}$ lists the measured FWHM of the main lobe of the ``dirty'' RM spread function 
(RMTF), and represents the actual Faraday resolution of our RM Synthesis spectra. 
\par
We quantify the statistical significance of $\phi$-component detections by imposing an amplitude threshold in the RM clean algorithm. Our target confidence is set at $\gtrsim 99\%$. To this end, we select 1000 off-source points in each of our image cubes, extract the $Q, U$ and $I$ fluxes in a similar fashion as described previously. We then apply our RM clean algorithm and for each science target image cube, we determine the maximum amplitude $|F|_{max}$ from the set of off-source positions, \black{above which the Probability$(|F|_{max,i}\geqslant |F|_{max}) \leqslant 0.001$. This gives us a $0.999$} confidence of individual RM components for each of our science targets. Finally, we combine this limit with the instrumental polarization leakage error estimate of $0.06$ \% by adding them in quadrature and using the result as the lower threshold for 
detection of Faraday components. 
Table \ref{rm-components} lists the thresholds and detected Faraday components for each source. We compared the RM Synthesis results with and without a weighting of $\sigma^{-2}_{qu}$ applied and found that the results do not differ. We thus apply no weighting in our final RM Synthesis analysis.
\par
We characterize Faraday complexity according to the second moment about the mean, $\sigma_{\phi}$, of the RM clean model spectrum \citep{Brown2011,Anderson2015} given by:

\begin{equation}
\sigma_{\phi} = \sqrt{\frac{\sum_{i}{(\phi_i - \mu_{\phi})^2|F(\phi_i)|}}{\sum_{i}{|F(\phi_i)|}}},
\end{equation}
with 
\begin{equation}
\mu_{\phi} = \frac{\sum_{i}{\phi_i|F(\phi_i)|}}{\sum_{i}{|F(\phi_i)|}}.
\label{eq:mu_phi}
\end{equation}
A Faraday simple spectrum will have $\sigma_{\phi} \approx 0$ rad\,m$^{-2}$ while the complex counterpart will have $\sigma_{\phi}$ very different from zero. We could not reliably resolve the complexity in individual components given that $\delta\phi^{\prime} \sim 100$ rad\,m$^{-2}$ but we classify a source spectrum as Faraday complex (denoted by C in Table \ref{rm-components}) or simple (denoted as S) according to  $\sigma_{\phi}$ calculated from all components that are detected above the noise. Table \ref{rm-components} summarizes the $\phi$-components we found near the RM clean threshold level for each source. The {\it major/main/prominent} (highest signal-to-noise ratio, S/N) components show some width manifesting as a number of $minor$ (low S/N) components separated by less than $\delta\phi^{\prime}$ from the main component. A Faraday emission component, with $\phi = \phi_i$, is called {\it main/most prominent} if $|F(\phi_i)| = $ maximum$(|F(\phi)|)$, and {\it minor} otherwise. The $\phi_i$ of minor components within $\delta\phi^{\prime}$ of each other are combined to give $\mu_{\phi}$, of detections as listed in Table \ref{rm-components}. The $1\times\sigma$ deviations of these minor components is reported as the error on $\mu_{\phi}$.
\begin{table*}
\centering
\caption{RM clean components for each of our sources. $|F|_{comp}$ is the peak of the Faraday spectrum derived during the RM clean algorithm, $\mu_{\phi}$ (equation \ref{eq:mu_phi}) represents the Faraday depth of each component, ${\sigma_{\phi}}$ is the second moment of the RM clean model  spectrum \citep{Brown2011,Anderson2015}, and the RM clean threshold. The column that follows, ``Class", shows classification of each source as Faraday complex (C) or Faraday simple (S) according to the value of ${\sigma_{\phi}}$. The last three columns show galactic coordinates ($l,b$) and also estimates of the smooth Galactic foreground Faraday depths, $\phi_{gal}$, at the locations of each our targets \citep{Oppermann2012,Oppermann2014}. }
\begin{tabular}{lcccccccc}
\hline\hline\\
[-1.25ex] 
Source& $|F|_{comp}$ & $\mu_{\phi}$   & ${\sigma_{\phi}}$ & Threshold &  Class & $l$  & $b$ & $\phi_{gal}$ \\
      & [$\%$]       & [rad\,m$^{-2}$]& [rad\,m$^{-2}$]   & [$\%$]    &        & [deg]&[deg]& [rad\,m$^{-2}$] \\
[1.25ex] 
\hline\\
[-1.25ex] 
J$0240-231$ & 1.62 & $8.0\pm 0.1$ & $37.7$ & 0.061  & C & 39.415 & -23.224 & 9.1 \\
				& 1.48 & $33.9\pm 1.4$   & &   & & &   &  \\
				& 1.29 & $-38.1\pm 4.8$  & &   & & &   &  \\
                & 0.69 & $-110.8\pm 2.0$  & &   & & &   &  \\
				& & & & & &   &  \\
J$0538-440$ & 0.45 & $91.2\pm 9.6$ & $40.8$ & 0.061 & C & 84.491 & -43.997 & 42.8 \\
				& 0.17 & $9.6\pm 1.4$  &  &    &  & &   &  \\
				&			&		 &  & & &   &  \\
NGC 1097 & 1.09 & $-21.5\pm 8.9$ & $120.2$ & 0.072 & C & 41.579 & -30.275 & 9.1 \\
				& 0.81 & $44.6\pm 3.5$ & &   & & &   &  \\
				& 0.47 & $147.3\pm 1.9$  & &  & & &   &  \\
				& 0.23 &$-520.5\pm 0.3$   & &  & & &   &  \\
				& 0.23 & $793.0\pm 0.3$  & &  & & &   &  \\
				& 0.22 & $619.9\pm 0.8$  & &  & & &   &  \\
				&	   &                  & &  & & &   &  \\
NGC 1808 & 0.43 & $-0.5\pm 1.3$  & $49.4$  & 0.065 & C & 76.926 & -37.513 & 35.3 \\
        		& 0.24 & $144.7\pm 0.7$   &  &  & & &   &  \\
        		& 0.12 & $-291.6\pm 0.1$   &  &  & & &   &  \\
				&            &       &    & & &   &  \\
PKS $1934-638$ & --         & --    & -- & 0.061 & -- & 65.146 & -63.713 & 44.7 \\
				&   		 &     &  &  &  \\
PKS B$0407-658$ & 0.15 & $8.0\pm 0.1$ & $0.0$  &  0.061 & S & 62.085 & -65.753 & 23.9 \\
\hline
\end{tabular}
\label{rm-components}
\end{table*}
\begin{figure*}
    \begin{subfigure}[t]{1.0\hsize}
    \centering
    \includegraphics[width=0.8\textwidth]{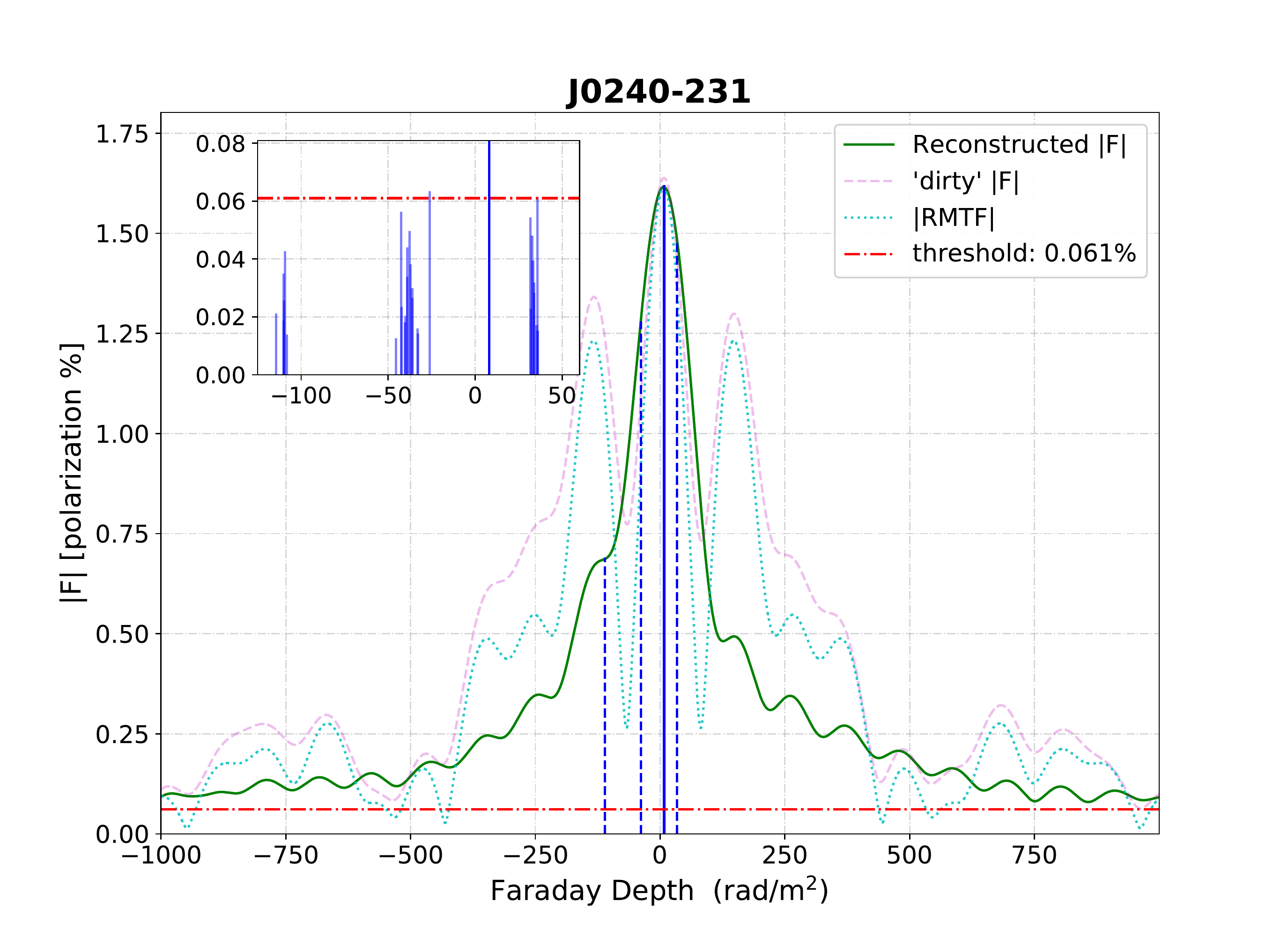}
    \subcaption{The clean Faraday spectrum, $|F(\phi)|$, of J$0240-231$ resulting from our RM clean algorithm. See main caption below.}
    \label{0240-rmclean}
    \end{subfigure}
    
    \begin{subfigure}[t]{1.0\hsize}
    \centering
    \includegraphics[width=0.8\textwidth]{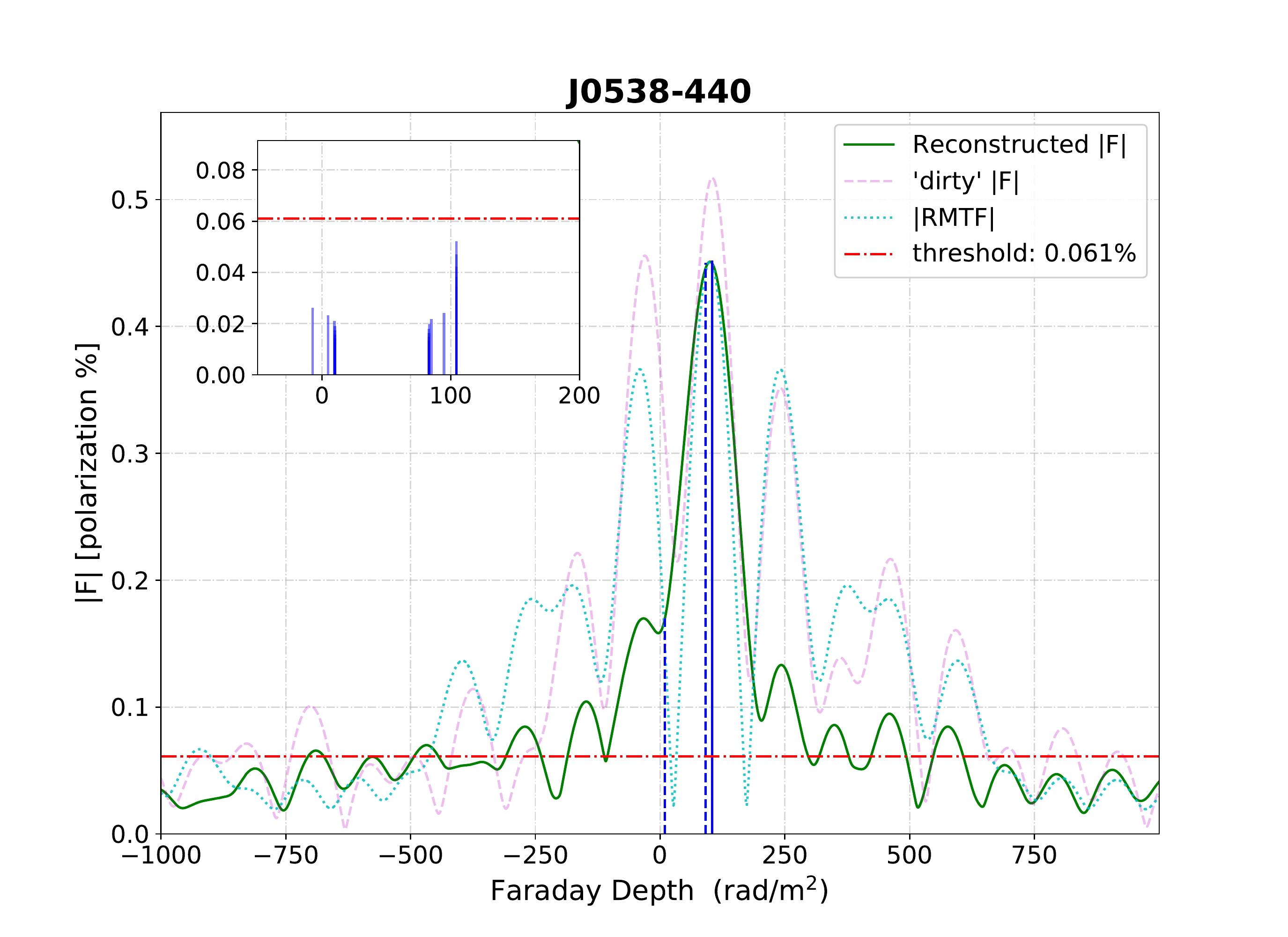}
    \subcaption{ The clean Faraday spectrum, $|F(\phi)|$, of J$0538-440$ resulting from our RM clean algorithm. See main caption below.}
    \label{0538-rmclean}
    \end{subfigure}
\end{figure*}

\begin{figure*}\ContinuedFloat
    \begin{subfigure}{1.0\hsize}
    \centering
    \includegraphics[width=0.8\textwidth]{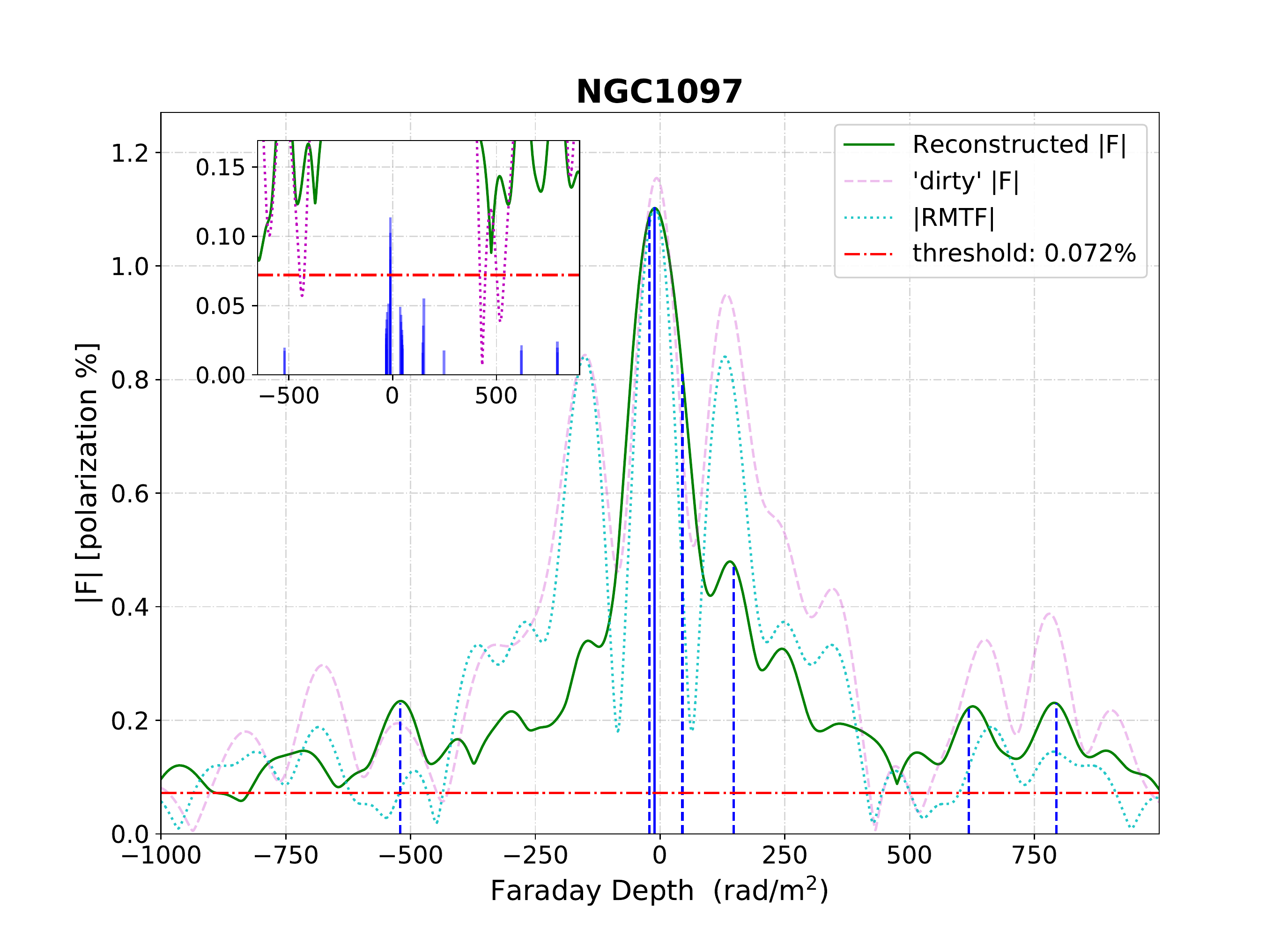}
    \subcaption{ The clean Faraday spectrum, $|F(\phi)|$ (green), of NGC 1097 resulting from our RM clean algorithm. See main caption below.}
    \label{NGC1097-rmclean}
    \end{subfigure}
    
    \begin{subfigure}{1.0\hsize}
    \centering
    \includegraphics[width=0.8\textwidth]{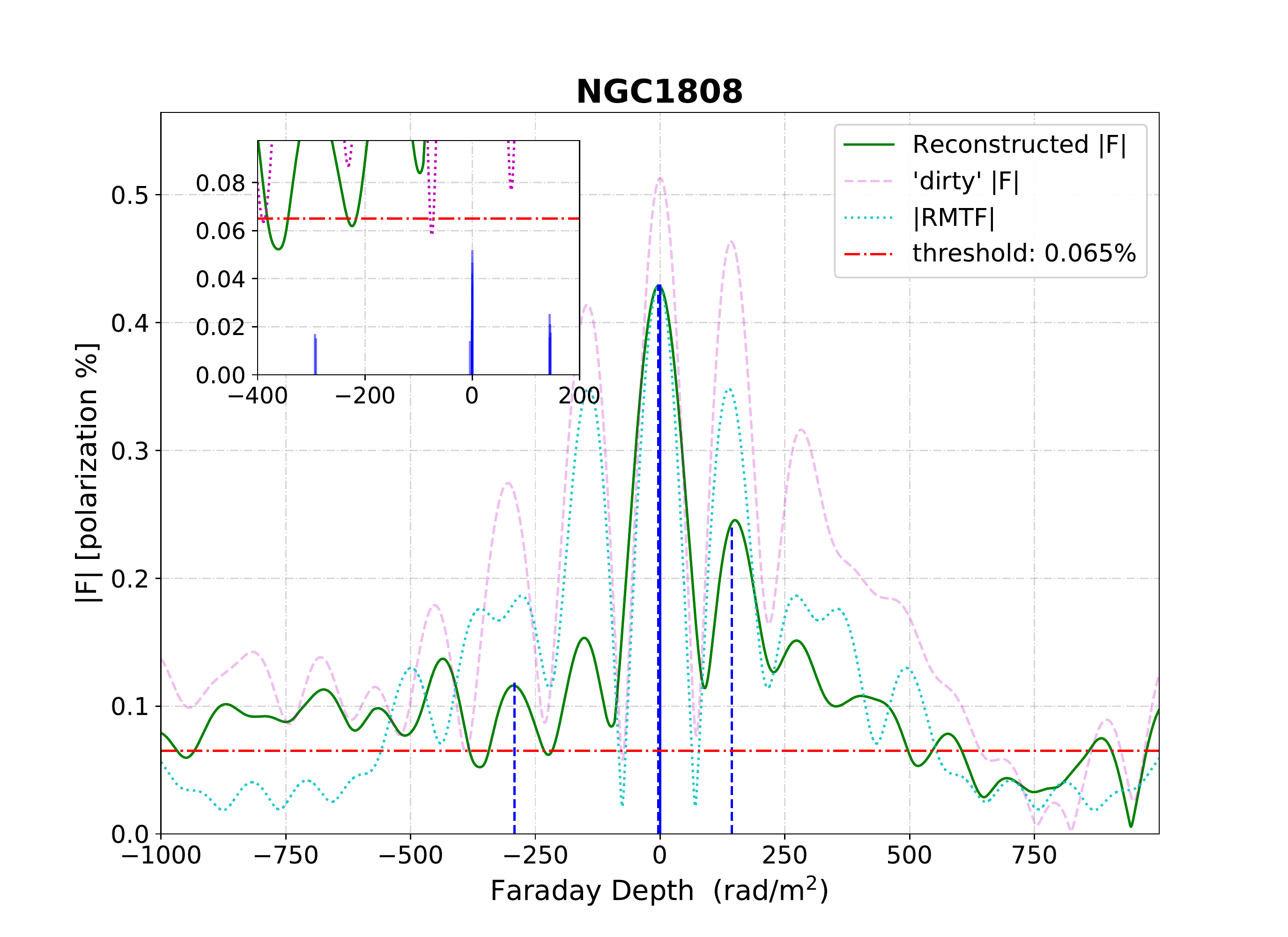}
    \subcaption{ The clean Faraday spectrum, $|F(\phi)|$, of NGC 1808 resulting from our RM clean algorithm. See main caption below. }
    \label{NGC1808-rmclean}
    \end{subfigure}
\end{figure*}
    
\begin{figure*}\ContinuedFloat
    \begin{subfigure}{.9\textwidth}
    \centering
    \includegraphics[width=0.8\textwidth]{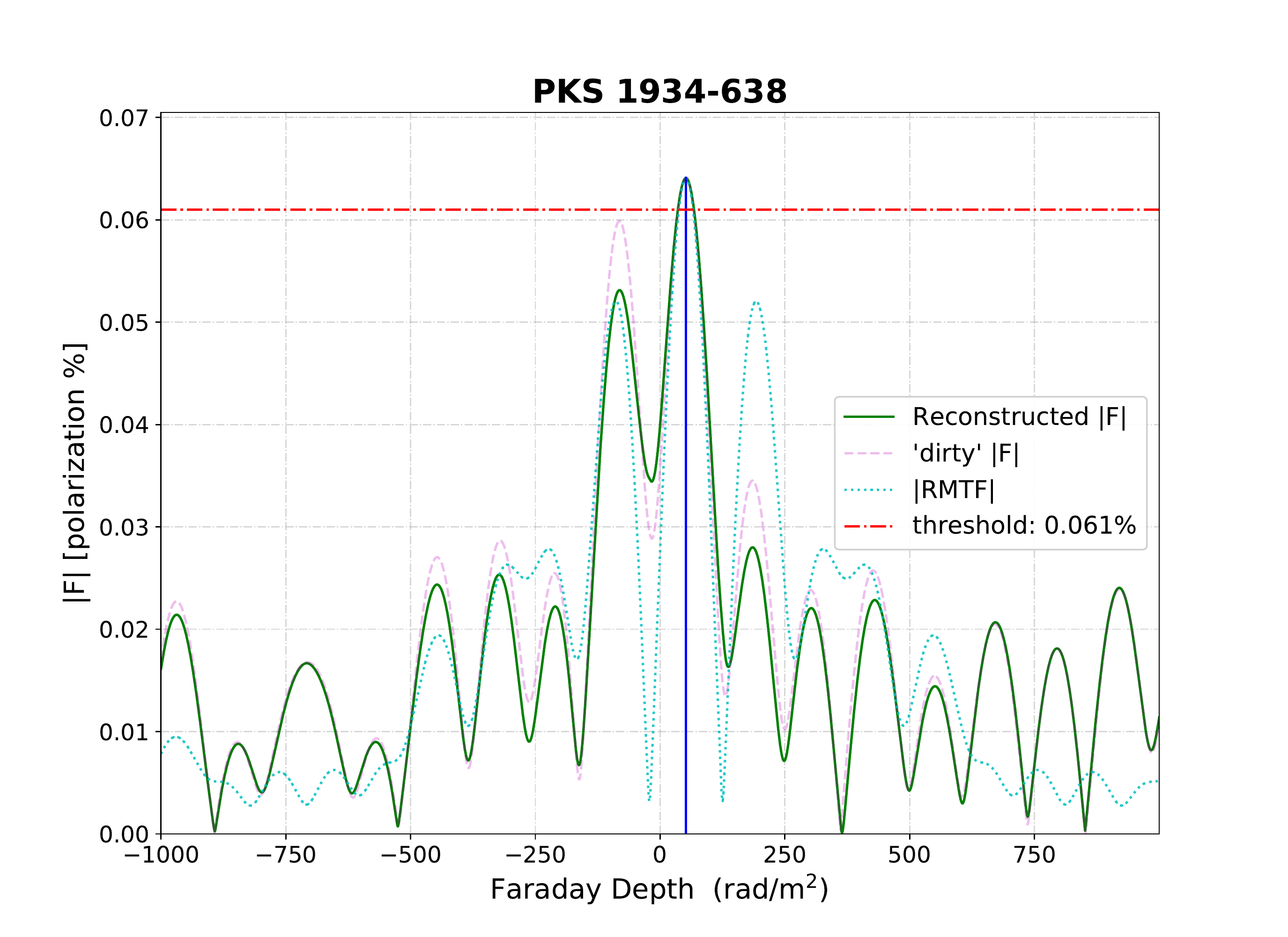}
    \subcaption{ The clean Faraday spectrum, $|F(\phi)|$, of PKS $1934-638$ resulting from our RM clean algorithm. See main caption below. }
    \label{pks1934-rmclean}
    \end{subfigure}

    \begin{subfigure}{.9\textwidth}
    \centering
    \includegraphics[width=0.8\textwidth]{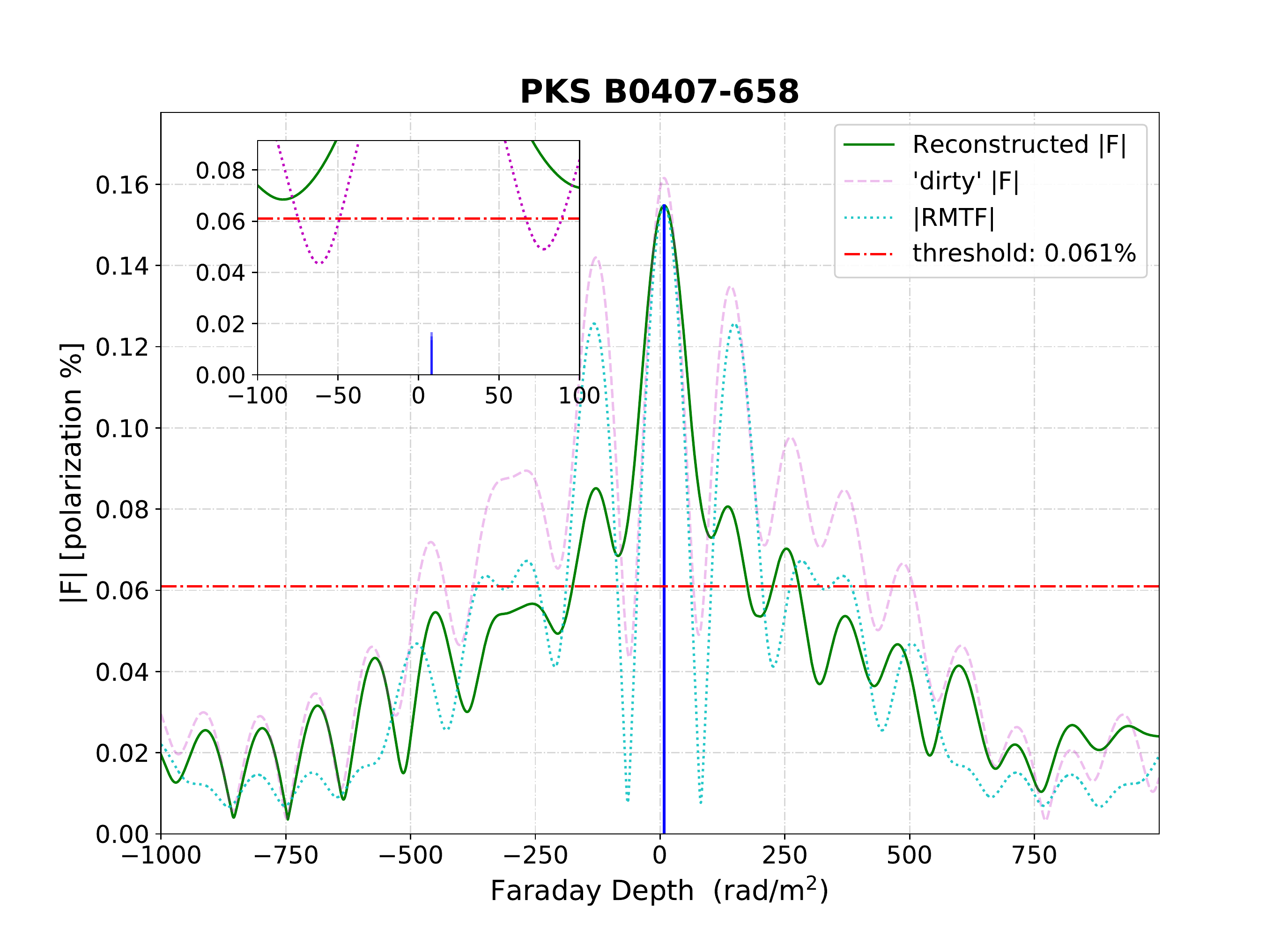}
    \subcaption{ The clean Faraday spectrum, $|F(\phi)|$, of PKS B$0407-658$ resulting from our RM clean algorithm. See main caption below. }
    \label{PKSB0407-rmclean}
    \end{subfigure}
\caption{Figures \ref{0240-rmclean} - \ref{PKSB0407-rmclean}: The clean Faraday spectrum, $|F(\phi)|$ resulting from our RM clean algorithm (green solid curve) with residuals added back in.  Top left: Zoom-in of the main plot showing deconvolved Faraday emission components (blue solid vertical lines show the peak $\gamma|F(\phi_i)|$ at the corresponding $\phi_i$ of each clean component in the vicinity of the brightest components) -- $\gamma$ is the loopgain parameter set to 0.1. The red dash-dotted horizontal line is the RM clean threshold as quoted in the legend. The dashed magenta curve is the ``dirty" Faraday spectrum while the dotted cyan curve is the RMTF scaled to the amplitude of the deconvolved $|F(\phi)|$ and with peak located at the Faraday depth corresponding to the peak of $|F(\phi)|$. Blue vertical dashed lines in the unzoomed plot display the value of $\mu_{\phi}$ of each detected Faraday component as quoted in Table \ref{rm-components}.}
\end{figure*}

Figures~\ref{pquX_fullband-ngc1097} - \ref{pquX_fullband-fluxcals} show the $\lambda^2$ dependence of $q$, $u$ and $p_o$ along with the model spectra reconstructed from the RM Synthesis clean components for each source. Figures~\ref{0240-rmclean} - \ref{PKSB0407-rmclean} show the ``dirty'' $F(\phi)$ for each source along with the RM Synthesis clean components. The strong side-lobes of the RM transfer function from the missing mid-band data set are visible in the dirty spectra. A ``cleaned'' $F(\phi)$ is also plotted, constructed by convolving the RM clean components with an idealized window function corresponding to full frequency coverage over the frequency range of the observations. The results for each source are discussed below.
\par
{\bf J$0240-231$}: This is the most polarized source in our sample with several Faraday components that are above the detection threshold -- see Figure \ref{0240-rmclean}. The most prominent component, as listed in Table \ref{rm-components}, is roughly consistent with the value of 14.4 $\pm$ 2.2 rad\,m$^{-2}$ measured by \cite{Taylor2009} based
on a simple two-point measure of polarization angle. Several Faraday components are seen in Figure \ref{0240-rmclean} and indicate that the fainter components are Faraday thick. Generally, the Faraday spectrum is complex with $\sigma_{\phi} = 37.7$ rad\,m$^{-2}$. 
\par
{\bf J$0538-440$}: This source shows two relatively thick components, each composed of a number of closely adjacent minor components (Figure \ref{0538-rmclean}). The peaks secondary in prominence to the two most prominent, seen as the lower level clean components in Figure \ref{0538-rmclean}, have been assumed to form part of the thick components. The source is classified as Faraday complex with $\sigma_{\phi}= 40.8$ rad\,m$^{-2}$.
\par
{\bf NGC 1097}: Displays the most complex Faraday depth profile amongst the two disk galaxy sources -- Figure \ref{NGC1097-rmclean}, with thick components at low $\phi$ and some faint components just above the threshold at large $|\phi| > 500$ rad\,m$^{-2}$ (particularly near $-520,\ 590,$ and $790$ rad\,m$^{-2}$). The most prominent component seems Faraday thick, with minor components spanning the range $\phi = -11$ to $-32$ rad\,m$^{-2}$. This source is classified as Faraday complex and displays the largest dispersion in Faraday space with $\sigma_{\phi} = 120.2$ rad\,m$^{-2}$. The component detected at $147.3\pm 1.9$ rad\,m$^{-2}$ most likely has some sidelobe emission from the first sidelobe, at $\phi = +129.0$ rad\,m$^{-2}$, but cannot be solely due to the sidelobe as the same location reflected about 0.0 rad\,m$^{-2}$ does not show the same detectable signal.
\par
{\bf NGC 1808}: Displays a Faraday spectrum with three Faraday-thin components -- Figure \ref{NGC1808-rmclean}. We detect three prominent $\phi$-components. The component with $\mu_{\phi} = 144.7\pm 0.7$ rad\,m$^{-2}$ is close to the first RMTF sidelobe located at $\phi = +140.5$ rad\,m$^{-2}$, however $|F(\phi)|$ is more than $3\times$ the detection threshold, and its presence is clearly seen as an asymmetry in the dirty Faraday spectrum. The component located at $\mu_{\phi} = -291.6\pm 0.1$ rad\,m$^{-2}$ also appears to coincide with the second RMTF sidelobe. The dirty Faraday spectrum hints at a component near the positive (i.e the RMTF in the $\phi > 0.0$ rad\,m$^{-2}$ region) second sidelobe but there are no detectable clean components in the region. The source is Faraday complex with $\sigma_{\phi} = 49.4$ rad\,m$^{-2}$.
\par
{\bf PKS $1934-638$}: This source is undetected in polarization. The peak emission in the Faraday spectrum is near $\phi = 50.0$ rad\,m$^{-2}$ with a peak $|F| \approx 0.061\%$ ($\approx 0.064\%$ when the residuals are added in), which is set as the instrumental polarization level. Figure \ref{pks1934-rmclean} displays the Faraday spectrum.
\par
{\bf PKS B$0407-658$}: The Faraday spectrum of this source displays a single Faraday-thin component and does not show any other components above the threshold (Figure \ref{PKSB0407-rmclean}. The lone component is detected at $\mu_{\phi} = 8.0 \pm 0.1$ rad\,m$^{-2}$ with a peak $|F| = 0.15\%$. 
\par

\subsubsection{Galactic Foreground and Ionospheric Effects}\label{Ionosphere-Subsect}
\black{Foreground RM contributions, such as from the ISM in the Galaxy and the Earth's ionosphere, can also be significant along the line of sight. This is especially so at frequencies below 1 GHz \citep[e.g.][]{Erickson2001}. However, most of the Galactic foreground emission is resolved out by an interferometer because of missing short spacings. There can be small scale structure in Stokes $Q$ and $U$ that the interferometer detects, and that can in principle become visible as a component in the Faraday depth spectrum. These are sometimes called Faraday Ghosts and are independent of position on the sky \citep[e.g.][]{Shukurov2003}.}
\par
\black{\cite{Oppermann2012} and \cite{Oppermann2015} have created a map of the smoothly varying Galactic Faraday component, $\phi_{gal}$, by utilising all-sky Faraday rotation measures of compact background sources from the literature. We list $\phi_{gal}$ at the locations of each of our targets in Table \ref{rm-components}. The effect of the Galactic ISM on an extragalactic source is mainly as a foreground Faraday-rotating screen. The correction for this effect is to subtract $\phi_{gal}$ from detected components. In this work, we note that this effect does exist and has possibly affected our measurements but we do not correct for it as it is not adequately constrained.}
\par
\black{In an effort to analyse Faraday complexity introduced by the ionosphere at the different observation epochs, we look for rotation in the polarization angle ($\chi$) as a function of time. This we do by producing full Stokes images for each scan of the phase calibrator for each epoch. We then restrict our analysis to those scans that have Stokes $V$ flux and $p_0$ that are below the leakage, and also Stokes I that is above $3\sigma$. We then analyse the resulting $q$, $u$ and $\chi$ as functions of scan i.e time. However, the resulting images prove to be much too noisy to lead to a conclusion on ionospheric effects.}
\par
\black{Typical Faraday rotation measure variations from the ionosphere are up to $\sim 1$ rad m$^{-2}$. Typical variations are of order one radian at 1m wavelength as $\chi(\lambda^2)\sim \lambda^2$. Variations of this order are seen at 327 MHz at the VLA \citep{Erickson2001}.  At our lowest frequency even a 90 degree change at 327 MHz scales to only about 6 degrees at 1300 MHz, and 2.5 degrees at 1900 MHz. This is small enough to ignore. Note also that it is only the time variation during an epoch that will affect the observations. A constant offset from the ionosphere will be accounted for by the calibration. In the absence of an absolute polarization calibrator, the solution is referenced to the mean feed position angle which is not expected to change as it is hardwired into the instrument. We therefore conclude that the expected ionospheric rotation is negligible and thus not corrected.}

\section{Discussion}\label{Discussion}

Our RM Synthesis analysis shows Faraday complexity, of varying degrees, in four of the six sources in our KAT 7 sample. The most polarized source, J$0240-231$, displays the most complex Faraday spectrum with several emission components (see Figure \ref{0240-rmclean} and Table \ref{rm-components}). The presence of Faraday thick components can be seen in the Faraday spectrum as well. Our analysis clearly resolves this complex spectrum both near the main peak and the smaller peaks. This source also shows the most rapidly changing polarized intensity that depolarizes toward larger wavelengths (see the bottom panel of Figure \ref{pquX_fullband-ngc1097}). The behaviour of $q(\lambda^2)$ and $u(\lambda^2)$ is also far from the simple sinusoidal behaviour, displayed by a single Faraday component, and this deviation is best illustrated in Figure \ref{pquX_fullband-ngc1097}. This plot resembles a circle of constant radius in the Faraday simple case, but we see quite a substantial deviation from this. Our RM clean model can however, reproduce the behaviour we see in the spectra. The $q(\lambda^2)$ and $u(\lambda^2)$ profiles have little scatter and are in good agreement with our RM clean model adding confidence to the derived $\phi$-components. \cite{Anderson2016} show that Faraday-complexity is common in highly polarized sources and our observation in the case of J$0240-231$ is in accord. \citet{Schnitzeler2019}, \black{however, fit one Faraday component to the J$0240-231$ spectrum through a QU-fitting analysis. The source was found to be significantly polarized at a reference frequency of 2.1 GHz with $p = 2.28\%$ suggesting agreement with the measurement made in this work at the high, 1.85 GHz, band.}
\par
Faraday complex sources are also found to more likely have steep spectra in the optical and radio  \citep[e.g.][]{Scarpa1997,Anderson2015}.  We find J$0240-231$ to have only a mildly steep total intensity spectrum of \myavg{\alpha}$_{1.85}^{1.35}=-0.562 \pm 0.006$. Our results for J$0240-231$ agree with previous findings that the source is significantly polarized with $p_{4.9GHz}=3.47\%$ \citep{Perley1982} and $p_{1.4GHz} = 0.8\%$ \citep{Taylor2009}. The object is a bright compact radio source \citep{Fanti1990,Fanti2000}, however, the double lobed structure reported by \cite{Dallacasa1998} may contribute to the observed Faraday complexity as radio lobes can be sites of complex Faraday structures \citep[e.g.][]{Sebokolodi2019,Banfield2019}.
\par
The case of J$0538-440$ is another where we find evidence of Faraday complexity, although not to the level of J$0240-231$ (contrast Figure \ref{0538-rmclean} with Figure \ref{0240-rmclean}). The behaviour of $q(\lambda^2)$ and $u(\lambda^2)$ profiles for this source (bottom panels of Figure \ref{pquX_fullband-ngc1808}) resemble that of a simple one-component Faraday spectrum with somewhat larger scatter than in J$0240-231$, indicating that perhaps higher sensitivity observations with a larger $\lambda^2$ coverage and thus Faraday depth resolution are needed to further explore the behaviour and further resolve the $\phi-$components. This source is highly polarized, $\sim 18\%$, in the optical \citep{Giommi1995,Scarpa1997} and has possibly suffered significant depolarization at L-band. Our RM clean implementation is able to model the behaviour well at the $\sim 2.5\sigma$ level imposed. \black{The J$0538-440$ total intensity spectrum is more time variable but appears flatter than in the case of J$0240-231$ with \myavg{\alpha}$_{1.85}^{1.35}= -0.098 \pm 0.003$ and a lower degree of polarization of less than $1\%$ at L-band.} \black{This is another source that is also in the \citet{Schnitzeler2019} catalog where it is found to be Faraday simple and highly polarized, $p=3.32\%$ (at $RM \approx 60 $ rad m$^{-2}$), at 2.1 GHz in contrast with $p=0.3 \pm 0.07\%$ measured at the high band in this work. The total intensity of this source is a highly time variable and we suspect that the significant discrepancy we observe is due to this variability as sources that are variable in total intensity may also be variable in polarization \citep{Anderson2019}. We observed this source in late July 2016 as compared to the 2011 observations of \citet{Schnitzeler2019}. The authors obtained their results from fitting simple Faraday components to the sources spectrum which differs significantly from our open ended approach in which our model can contain as many (clean) components as is necessary to fit the data. With QU-fitting, a much more constrained model is fit to the data while minimizing errors. We do not apply constraints to this work. Another contributing factor to observed discrepancies is the difference in frequency bands, L and S-band. At L-band, it is typical that polarization is lower than at S-band, especially for compact sources.}
\par
NGC 1097 displays the most complex Faraday spectrum in the two disk galaxies. The Faraday spectrum of the source shows the largest spread in $\phi$ (see Figure \ref{NGC1097-rmclean}). Low S/N emission at comparatively large $\phi$ could be contaminated by sidelobes of the RMTF but may also be due to small regions of highly magnetised media which may emit at large Faraday depths. These small regions may coincide with highly turbulent regions near the core of a galaxy \citep[e.g][]{Pasetto2018}. The low S/N of these components may also point to depolarization that may be due to turbulence of the magnetised plasma in the bar and circumnuclear ring regions of the galaxy and/or frequency dependent depolarization that diminishes the polarized signal at L-band. Higher resolution measurements by \cite{Beck2005} do not show evidence of high rotation measure regions in NGC 1097 (the authors report $\phi$ in the range -200 to +200 rad\,m$^{-2}$ from 50 MHz bandwidth observations at 4.86 and 8.46 GHz) and so we may be detecting a noise contribution to the signal that is unaccounted for. More sensitive observations with a wider bandwidth coverage will be required to confirm these faint components. Polarimetry at higher frequency may also be able reveal such components in the absence of internal depolarization effects \citep[e.g.][]{Anderson2016}.
\par
NGC 1808 is also classified as Faraday complex in our RM clean analysis with the detection of three Faraday-thin components -- see Figure \ref{NGC1808-rmclean}. The complexity classification in this disk galaxy is due to the fairly large spread of Faraday depths found. Our RM-clean implementation shows subtle structure of the polarized spectra of the galaxy while the data shows a low S/N and significant scatter at the wavelengths observed. A larger $\lambda^2$ coverage with higher sensitivity is required as we will be able to better constrain the polarized spectra of such Faraday complex and low polarized S/N sources \citep[e.g][]{OSullivan2012,Anderson2015,Anderson2016}. 
\par
\black{The apparent spectral indices we derive suggest agreement with those expected for spiral galaxies; $\alpha=-0.8 $ to $-1.0$, \citep{Beck2015}. We find \myavg{\alpha}$_{1.85}^{1.35} = -0.815 \pm 0.011$ for NGC 1808 and \myavg{\alpha}$_{1.85}^{1.35} = -1.060 \pm 0.020$ for NGC 1097. NGC 1808 is weakly polarized, with $p<1\%$ in accordance with models of the integrated polarization of galaxies at 1.4 GHz \citep{Stil2009,Sun2012}. The source also shows evidence of possible polarized emission components emitted at high Faraday depths as suggested by low level (undetected) large $\phi$ emission in the deconvolved Faraday spectrum.}
\par
PKS $1934-638$ \black{is not detected in polarization}. This source shows polarization levels that are consistent with noise across the L-band. \cite{Eichendorf1980} quote polarization levels that are of order double what we have found with observational errors typically being in the range of 1 to 2 percent in 1965 and lowering to 0.5 to 1 percent in 1974. We interpret their result as upper limits. Our results are comparable to the earlier data (Table \ref{Eichendorf_1980Table}) showing that PKS $1934-638$ is undetected in polarization.
\par
PKS B$0407-658$, however, is found to have one detected component -- Figure \ref{PKSB0407-rmclean}. The $q(\lambda^2)$, $u(\lambda^2)$, and $p(\lambda^2)$ profiles of this source are not as featureless as in PKS $1934-638$ and may require a larger $\lambda^2$ coverage to further resolve the Faraday spectrum. More sensitive MeerKAT commissioning observations show that PKS B$0407-658$ is polarized at $\sim 0.03 - 0.05\%$ levels in the L-band \citep{Hugo2018}.
\par
\black{There exists no known relation between Faraday complexity and morphological complexity at the angular scales probed here \citep[e.g.][]{Anderson2015}} so even the most complex sources may be much simpler in morphology. More sensitive and higher resolution broad-band spectropolarimetry studies are needed to further resolve Faraday components in Faraday depth, across wide bandwidths, at higher frequencies, and also spatially with high S/N as will be the case with the next generation of radio telescopes such as MeerKAT.

\section{Summary and conclusions}\label{Summary+conclusions}
We have explored the properties of broadband linear polarization in two galaxies and four AGN powered sources in direct pointing mode with the KAT 7 array. The observations were made at the KAT 7 low band ($\nu_c \sim 1350$ MHz) and high band ($\nu_c \sim 1850$ MHz) with the mid band ($\nu_c \sim 1550$ MHz) discarded due to RFI. The synthesized beam at the low band was $\sim 4.9^{\prime}$ and the same at the high band was $\sim 3.9^{\prime}$. We detect linear polarization down to $\sim 2$ mJy/bm. We employed the open-ended method of RM Synthesis and RM clean \citep{Burn1966,Brentjens2005,Heald2009} which enabled us to resolve the Faraday depth components down to a Faraday depth resolution $\delta\phi^{\prime}\sim 100$ rad\,m$^{-2}$ along our line of sight. Our conclusions are summarized as follows:

\begin{itemize}

\item[({\bf i})] The total intensity spectra that we were able to measure are in general agreement with previous studies. 

\item[({\bf ii})] Our spectropolarimetric \black{analysis} shows Faraday simple, PKS B$0407-658$, \black{unpolarized PKS $1934-638$}, and Faraday complex sources \black{-- J$0538-440$, J$0240-231$, NGC 1097 and NGC 1808}. The most polarized objects display more complex Faraday spectra which is indicative of multiple emitting and/or diminishing components at different Faraday depths along the line of sight.

\item[({\bf iii})] \black{Our RM Synthesis analysis did not detect Faraday complexity in the least polarized source (PKS B$0407-658$) but such sources may still have some level of underlying Faraday complexity that is beyond detection}. The two disk galaxy sources are classified as complex and have been found to have emission components over larger Faraday depth ranges -- suggesting complex magneto-ionic emission environments along the line of sight.

\item[({\bf iv})] The BL Lac source, J$0538-440$, and QSO, J$0240-231$, are the most highly polarized in the literature and most Faraday complex of our direct pointing sources. Our measurements support earlier findings that Faraday complexity is expected in most polarized sources. The lower Faraday complexity found in the other less polarized sources may require more sensitive wide band and thus higher $\phi$ resolution studies to further tease out the properties that may be beyond our current data. Low polarization sources may also have a significant degree of spectropolarimetric complexity that is beyond detection in studies with similar sensitivity and resolution limitations as ours. \black{}.

\item[({\bf v})] We determine the on-axis instrumental polarization level in KAT 7 to have an upper limit of $0.06\%$. This is based on the polarization level measured for the hitherto unpolarized GPS source PKS $1934-638$ spectrum.

\item[\black{({\bf vi})}] \black{Some BL Lac sources, such as J$0538-440$, are particularly variable with time suggesting more complexity which should be accounted for when such sources are used as calibrators especially over broad bandwidths}

\end{itemize}

\section*{Acknowledgements}

This work is funded by the South African National Research Foundation (SA NRF) and the South African Radio Astronomy Observatory (SARAO), formerly known as the Square Kilometre Array South Africa (SKA SA) project, through their doctoral student scholarship. The analysis is done on the Inter-University Institute for Data Intensive Astronomy (IDIA) cloud computing facility.

\section*{Data availability}
The data underlying this article can be shared upon reasonable request to the corresponding author.
\bibliographystyle{mn2e}
\bibliography{ref_lib}
\label{lastpage}
\end{document}